\documentclass[12pt,preprint2,natbib,iop]{emulateapj} 
\usepackage{graphicx,epsfig,amsmath,multirow,hyperref,breakurl} 

\lefthead{Maseda et al.}
\righthead{Kinematics and Metallicites of Extreme Emission Line Galaxies}
\slugcomment{Version: 16 June, 2014}
\begin{document}

\newcommand{\kms}{\>{\rm km}\,{\rm s}^{-1}}
\newcommand{\reff}{r_{\rm{eff}}}
\newcommand{\msol}{M_{\odot}}
\newcommand{\zsol}{Z_{\odot}}
\newcommand{\inverse}[1]{{#1}^{-1}}
\newcommand{\invvar}{\inverse{C}}
\newcommand{\dd}{{\rm d}}

\title{The Nature of Extreme Emission Line Galaxies at {\small z}=1-2: Kinematics
and Metallicities from Near-Infrared Spectroscopy\altaffilmark{*}} 

\author{Michael V. Maseda\altaffilmark{1}, Arjen van der Wel\altaffilmark{1}, Hans-Walter Rix\altaffilmark{1},
Elisabete da Cunha\altaffilmark{1}, Camilla Pacifici\altaffilmark{2},  Ivelina Momcheva\altaffilmark{3}, Gabriel B. Brammer\altaffilmark{4}, Sharon E. Meidt\altaffilmark{1}, Marijn Franx\altaffilmark{5}, Pieter van Dokkum\altaffilmark{3}, Mattia Fumagalli\altaffilmark{5}, Eric F. Bell\altaffilmark{6}, Henry C. Ferguson\altaffilmark{4}, Natascha M. F\"orster-Schreiber\altaffilmark{7}, Anton M. Koekemoer\altaffilmark{4}, David C. Koo\altaffilmark{8}, Britt F. Lundgren\altaffilmark{9}, Danilo Marchesini\altaffilmark{10}, Erica J. Nelson\altaffilmark{3}, Shannon G. Patel\altaffilmark{11}, Rosalind E. Skelton\altaffilmark{12}, Amber N. Straughn\altaffilmark{13}, Jonathan R. Trump\altaffilmark{14}, Katherine E. Whitaker\altaffilmark{13}}

\affil{$^1$ Max-Planck-Institut f\"ur Astronomie, K\"onigstuhl 17, D-69117
Heidelberg, Germany; email:maseda@mpia.de}
\affil{$^2$ Yonsei University Observatory, Yonsei University, Seoul 120-749, Korea}
\affil{$^3$ Department of Astronomy, Yale University, New Haven, CT 06520, USA}
\affil{$^4$ Space Telescope Science Institute, 3700 San Martin Drive, Baltimore, MD 21218, USA}
\affil{$^5$ Leiden Observatory, Leiden University, Leiden, The Netherlands}
\affil{$^6$ Department of Astronomy, University of Michigan, 500 Church Street, Ann Arbor, MI 48109, USA}
\affil{$^{7}$ Max-Planck-Institut f\"ur extraterrestrische Physik, Giessenbachstrasse 1, D-85748 Garching, Germany}
\affil{$^8$ UCO/Lick Observatory and Department of Astronomy and Astrophysics, University of California Santa Cruz, 1156 High Street, Santa Cruz, CA 95064, USA}
\affil{$^9$ Department of Astronomy, University of Wisconsin, 475 N. Charter Street, Madison, WI 53706, USA}
\affil{$^{10}$ Physics and Astronomy Department, Tufts University, Robinson Hall, Room 257, Medford, MA 02155, USA}
\affil{$^{11}$ Carnegie Observatories, 813 Santa Barbara Street, Pasadena, CA 91101, USA}
\affil{$^{12}$ South African Astronomical Observatory, P.O. Box 9, Observatory 7935, Cape Town, South Africa}
\affil{$^{13}$ Astrophysics Science Division, Goddard Space Flight Center, Code 665, Greenbelt, MD 20771, USA}
\affil{$^{14}$ Department of Astronomy and Astrophysics, The Pennsylvania State University, Davey Lab, University Park, PA 16802, USA}
\altaffiltext{*}{This work is based on observations taken by the 3D-HST Treasury Program and the CANDELS Multi-Cycle Treasury Program with the NASA/ESA HST, which is operated by the Association of Universities for Research in Astronomy, Inc., under NASA contract NAS5-26555.  X-Shooter observations were performed at the European Southern Observatory, Chile, Program 089.B-0236(A).}
\begin{abstract}

We present near-infrared spectroscopy of a sample of 22 Extreme Emission Line Galaxies at redshifts $1.3 < z < 2.3$, confirming that these are low-mass ($M_{\star} = 10^8 - 10^9 \msol$) galaxies undergoing intense starburst episodes ($M_{\star}/SFR \sim$ 10 $-$ 100 Myr). The sample is selected by [O III] or $H\alpha$ emission line flux and equivalent width using near-infrared grism spectroscopy from the 3D-HST survey. High-resolution NIR spectroscopy is obtained with LBT/LUCI and VLT/X-SHOOTER. The [O III]/$H\beta$ line ratio is high ($\gtrsim 5$) and [N II]/H$\alpha$ is always significantly below unity, which suggests a low gas-phase metallicity.  We are able to determine gas-phase metallicities for 7 of our objects using various strong-line methods, with values in the range 0.05 $-$ 0.30 $\zsol$ and with a median of 0.15 $\zsol$; for 3 of these objects we detect [O III]$\lambda$4363 which allows for a direct constraint on the metallicity.  The velocity dispersion, as measured from the nebular emission lines, is typically $\sim 50 \kms$.  Combined with the observed star-forming activity, the Jeans and Toomre stability criteria imply that the gas fraction must be large ($f_{gas} \gtrsim$ 2/3), consistent with the difference between our dynamical and stellar mass estimates. The implied gas depletion time scale (several hundred Myr) is substantially longer than the inferred mass-weighted ages ($\sim$50 Myr), which further supports the emerging picture that most stars in low-mass galaxies form in short, intense bursts of star formation.

\end{abstract}

\section{INTRODUCTION} 
The life cycles of dwarf galaxies with present-day stellar masses $\lesssim10^{9}~\msol$ illustrate many challenges to our current understanding of galaxy formation and evolution. While we see that starbursts do not play an important role in the global star-formation at the present epoch \citep{lee}, it is likely that the star formation histories of dwarf galaxies are complex and varied \citep{mateo98} and that their typical star-formation rates were higher in the past \cite[e.g.][]{gallagher}.  That star formation in low-mass galaxies may be very burst-like is predicted by  hydrodynamical simulations \cite[e.g.,][]{pelupessy,stinson,shensims}.  In general, star formation appears to be regulated by stellar feedback in the form of supernovae and winds that heat and deplete the central reservoirs of cold gas required for continued star formation.  In simulations of lower mass systems, feedback is predicted to eject gas out of the galaxy and into the halo, resulting in an episodic star formation history 
across the entire galaxy \citep{stinson}. Repeated bursts have been cited as the 
driving 
force behind intense feedback mechanisms that can change the dynamical profile of the systems by driving baryons out of the center of the halo on short time scales, which also displaces the dark matter from the center and creates a cored dark matter profile \citep{n96,pontzen,governato,zolotov,amorisco}, potentially addressing one of the principal challenges to the standard $\Lambda$CDM cosmology.

Until recently, however, the progenitors of contemporary dwarf galaxies at high-redshift could not be studied directly.  ``Archaeological'' studies of resolved stellar populations in local dwarfs \cite[e.g.,][]{grebel, weisz} have confirmed that their star-formation histories are indeed not smooth, and have also found that a large fraction of their stars formed at early epochs ($z > 1$). For such old stellar populations ($>7$ Gyr), the age resolution achieved by the archaeological approach is insufficient to distinguish star formation events with durations of $10^7$ and $10^8$ years, leaving the actual ``burstiness'' of the star formation history unconstrained.

Look-back studies, directly observing the star formation activity in very low-mass systems at $z > 1$, were impractical until recently.  But, near-infrared spectroscopy and deep imaging from the Wide-field Camera 3 (WFC3) on the \textit{Hubble Space Telescope} (HST) has provided a new opportunity. An abundant population of galaxies at $z\sim 1.7$ with extremely high equivalent widths (EWs) was found by \citet{vdw} using data from the Cosmic Assembly Near-infrared DEep Legacy Survey\footnote{\url{http://candels.ucolick.org/}}  \cite[CANDELS,][]{candels1,candels2}.  They find a large number of objects with unusual $I-J$ and $J-H$ colors which imply the presence of a bright emission line in the $J$-band, likely to be [O III] after considering other photometric constraints.  Slitless grism spectroscopy confirms the redshifts and emission-line interpretation of four of these objects.  These are the high-EW tail of the emission line galaxies found in \citet{straughn} via slitless grism spectroscopy at the HST, a 
tail which is also probed by \citet{atek}.  
Other objects with similarly low masses and metallicities at these redshifts have been discovered in \citeauthor{gb2} (2012b; $EW_{[O III],rest}$=1499 \AA) and \citeauthor{vdw13} (2013; $EW_{[O III],rest}$=1200 \AA) assisted by strong gravitational lensing, in \citeauthor{masters} (2014; $EW_{[O III],rest}$=154 \AA), in \citeauthor{erb10} (2010; $EW_{[O III],rest}$=285 \AA), and in \citeauthor{maseda} (2013; discussed further below).  All of these objects are emission-line dominated systems with low metallicities and high equivalent widths, the so-called ``Extreme Emission Line Galaxies'' (EELGs).  

These systems are likely the high-EW tail of the distribution of high-redshift dwarf galaxies, and resemble the class of blue compact dwarf galaxies \cite[BCDs,][]{sargent} observed locally in several ways: low masses, high SFRs relative to their mass, and strong emission lines. However, the EELGs are indeed ``extreme,'' with sSFR values an order of magnitude higher than the BCDs, similar to the strong [O III] emitters (``green peas'') discovered photometrically in the SDSS by \citet{greenpea}, as well as spectroscopically by \citet{amorin} and \citet{izotovSDSS}.  As suggested by those authors, the star-formation mode exhibited in the ``green peas'' is likely a relic of a mode that was much more prevalent in the earlier Universe: their comoving number density is one to two orders of magnitude lower than the value of $3.7\times10^{-4}~Mpc^{-3}$ for EELGs at $z\sim 1.7$ \citep{vdw}.  

The implication of strong emission lines and an extremely faint continuum are that these systems have low masses and are undergoing an intense burst of star formation.  Equivalent widths $>$ 100~\AA$ $ and stellar masses of $10^8-10^9 ~\msol$ imply specific star formation rates (sSFR) in excess of 10 Gyr$^{-1}$, which is more than an order of magnitude higher than star-forming systems of equivalent masses at lower redshifts \citep{karim}.  These low stellar and dynamical masses are confirmed in \citet{maseda}, who also rule-out significant contributions to the total mass from older stellar populations for objects with restframe [O III]$\lambda5007$ equivalent widths $> 500$ \AA$ $ and intimate that the bursts are intense and have low metallicities. The implied star formation rates and masses have only been reproduced recently in hydrodynamical simulations \cite[e.g. in][]{shensims}.  

Although these previous observational studies have placed constraints on various quantities, many uncertainties remain.  In the case of \citet{maseda} and \citet{vdw}, all of the observed [O III], H$\beta$, and H$\alpha$ emission is attributed to star formation and not AGN.  In \citet{vdw}, upper-limits are placed on the black hole masses from the UV-continuum slopes, but their starbursting nature is merely plausible given the lack of knowledge about low-metallicity AGN \citep{izotovAGN, kewley13}.  Low metallicies are simply inferred from the consistency of the SED fits using low-Z (0.2 $\zsol$) templates with the observed photometry.  

These objects represent a field of growing importance, both for studies of low-mass dwarf galaxies and for the highest-z galaxies observed.  Depending on the strength, duration, and initial mass of these bursts, the descendants could display a wide range of masses, from present-day $\sim 10^9~\msol$ dwarf galaxies to potentially Milky Way-like systems, particularly if merging is important.  In any case, these galaxies are also significant contaminants in searches for higher-z dropout sources \citep{atek}, as in the case of \textit{UDFj-30546284} \citep{ellis,gb13,bouwens}.  As mentioned in \citet{atek} and \citet{coe}, great care must be taken in the interpretation of high-z candidates, as EELGs can potentially reproduce their observed colors (although the implied EWs in the latter case exceeded those that can be produced from star formation alone).

Some issues still remain in our understanding of such systems, such as stringent limits to the low masses and metallicities, including those with EWs $<$ 500 \AA, and the starburst origin of their strong emission lines.  Here we 
combine both high- and low-resolution near-IR spectroscopy with broadband 
photometry.  With the low-resolution spectra, we select candidates for follow-up high-resolution spectroscopy and obtain emission line fluxes.  The high-resolution spectra constrain various emission line ratios, some of which are useful diagnostics of AGN activity, as well as line widths, which are themselves a probe of the dynamical masses of the systems.  This provides strong evidence for their low masses and low metallicities, as well as confirming their starbursting nature.  Sophisticated modeling of the broadband SEDs  constrains the stellar masses and ages, as well as providing information on the dust content and metallicities. Together, this tells us about the strength and duration of the star-forming event.

The remainder of this paper is organized as follows.  In $\S$ 2 we describe the near-infrared spectroscopy and multi-band photometry used in the initial candidate selection process and the subsequent follow-up observations.  $\S$ 3 presents the results of the spectroscopic study, including emission-line widths, physical sizes, and masses.  In $\S$ 4 we confirm their low metallicities and rule out AGN as a significant source of contamination.  $\S$ 5 presents the implications of this work for the gas content of these systems, and $\S$ 6 summarizes our findings and puts the results into the overall context of the formation history of dwarf galaxies.  We adopt a flat
$\Lambda$CDM cosmology with $\Omega_m=0.3$ and H$_0=70 ~$km s$^{-1}$
Mpc$^{-1}$ and a \citet{chabrier03} IMF throughout.

\section{DATA}

\subsection{Candidate Selection}
\label{sec:sel}
In order to search for and investigate these ``starbursting dwarf galaxies,'' we take a multi-faceted approach.  Our preliminary search utilizes data from the 3D-HST survey\footnote{\url{http://3dhst.research.yale.edu/Home.html}} \citep{vd,gb}, a near-infrared spectroscopic Treasury program utilizing the WFC3.  This program provides WFC3/IR primary and Advanced Camera for Surveys (ACS) parallel imaging and grism spectroscopy over approximately three-quarters (625 square arcminutes) of the CANDELS fields.  The main source of spectroscopic data comes from the WFC3 G141 grism, with an effective wavelength coverage of 1.1 to 1.65 $\mu$m.

\begin{figure}

 \includegraphics[width=0.475\textwidth]{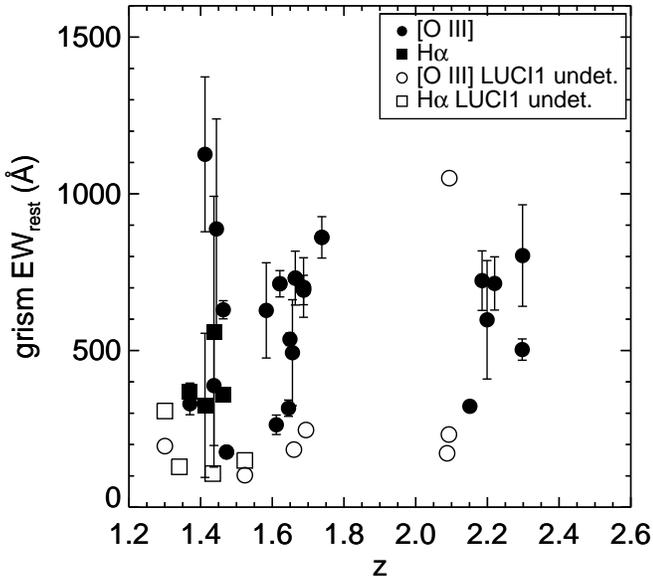}
\caption{Restframe equivalent widths as a function of redshift for our entire sample.  Circles represent equivalent widths from [O III]$_{5007}$ and squares represent equivalent widths from H$\alpha$.  The equivalent widths were determined from photometry and grism spectroscopy, see Section \ref{sec:sel}.  For objects in which both [O III] and H$\alpha$ are visible in the grism spectrum, both equivalent widths are plotted.  Open symbols show objects without LUCI1 line detections, due to intrinsic faintness or skyline contamination.  Our emission-line detection rate is close to 100\% for objects with EW $>$ 300 \AA.}
\label{fig:ewz}
\end{figure}

The grism data allow us to select and confirm strong line emitters spectroscopically.  Photometric cuts, such as the $iJH$ cut of \citet{vdw} and a similar $ViJH$ selection (excess in $H$ compared to a blue continuum, all from CANDELS), are used to preselect objects with strong features in their SEDs.  The G141 grism data are reduced according to \citet{gb} for the UKIDSS Ultra-Deep Survey (UDS), GOODS-South (GOODS-S), and COSMOS fields, and are then used to confirm bright lines with little or no associated continuum.  While we find numerous examples of these objects at $z > 1$ (Maseda et al. in prep.), we focus here on objects where the emission lines do not fall in the wavelength range between the $J$-, $H$-, and $K$-bands, enhancing the chances for detectability from the ground. The low-resolution WFC3 grism data ($R\sim 130$) provide redshift information such that targets can be selected with [O III] in the redshift range  1.15 $< z <$ 2.40 and H$\alpha$ in the redshift range 0.64 $< z <$ 1.59 to $\delta$z/z $\sim$ 0.005.  However, our photometric preselection relies on flux excesses such that we mostly see objects at 1.3 $\lesssim z \lesssim$ 1.8 and 2.1 $\lesssim z \lesssim$ 2.3.  Note that we do not resolve the continuum in our ground-based observations (discussed presently), so all EW values are calculated from the grism spectra directly.  In total, we select 31 objects for ground-based spectroscopic observations using this method.  An additional five candidates are taken from the sample of \citet{vdw} and one from \citet{straughn11}.  Their equivalent widths as a function of redshift are shown in Figure \ref{fig:ewz}.  Non-detections are due to either intrinsic faintness in the lines (we are sensitive to line flux in our ground-based spectra, not EW) or contamination from OH-skylines.

\subsection{LBT/LUCI1 Spectroscopy}

We observed our grism-selected sample with the LUCI1 multi-object spectrograph \citep{seifert} on the 8.4 m \textit{Large Binocular Telescope} (LBT).  We use LUCI1 in MOS mode, splitting our 31 candidates between four masks, during April 2012 (two masks in the COSMOS field), October 2012 (one mask in the UDS field), and March 2013 (an additional mask in the COSMOS field).  Approximately two hours are spent observing in each of the $J$- and $H$-bands for the first two COSMOS masks (A and B) using 1$''$ slits, two hours in each of the $H$- and $K$-bands for the UDS mask (C) with 0.6$''$ slits, and three total hours on the final COSMOS mask (D) in the $J$- and $H$-bands using 0.6$''$ slits.  All data are taken using the high-resolution $210\_zJHK$ grating ($R_J=8460$, $R_H=7838$, $R_K=6687$). The exposures are dithered by 3$''$ and are of varying durations, depending on the band: $J$-band data using 600s exposures, $H$-band data with 300s exposures, and $K$-band data with 120s exposures.  The shorter 
integrations in the $H$- and $K$-bands lead to lower signal-to-noise (S/N) due to additional readnoise, but are necessary so as not to saturate the detector in the regions with the brightest OH sky lines.  Seeing was generally good (between 0.5$''$ and 1$''$ in the optical) during the COSMOS observations, with good transparency.  For the UDS observations, seeing was generally better than 1$''$.  Table \ref{tab:obs} details the observations of individual objects, including IDs (from the CANDELS catalogs, discussed in Section \ref{sec:sed}), coordinates, and the main line detections.
\begin{deluxetable}{lcccc}
\tabletypesize{\scriptsize} 
\tablecaption{Summary of Near-IR Observations\label{tab:obs}}
\tablewidth{0pt}
\tablehead{ \colhead{ID} & \colhead{RA} & \colhead{Dec} & \colhead{Mask} & \colhead{Observed} \\
    & (deg) & (deg) & &Lines}\\
    \startdata
GOODS-S-7892&53.17194 &-27.75915&...&[O III], H$\alpha$\\
GOODS-S-43693&53.07129&-27.70580&...&H$\alpha$\\
GOODS-S-43928&53.05158&-27.70476&...&[O III]\\
UDS-6195&34.42648&-5.25577&...&[O III], H$\alpha$\\
UDS-6377&34.42857&-5.25532&...&[O III], H$\alpha$\\
UDS-7665&34.39076 &-5.25080 &C&[O III]\tablenotemark{a}\\
UDS-10138&34.42336 &-5.24226&C&[O III]\tablenotemark{a}\\
UDS-12435&34.41087 &-5.23481&C&H$\alpha$\tablenotemark{a}\\
UDS-12539&34.47389&-5.23423&...&[O III], H$\alpha$\\
UDS-12920&34.39870 &-5.23320&C&...\\
UDS-15319&34.40516& -5.22493&C&...\\
UDS-19167&34.43140 &-5.21212&C&[O III]\\
UDS-24154&34.39137 &-5.19531&C&[O III], H$\alpha$\\    
COSMOS-8509&150.09837 &2.26596 &B&...\\    
COSMOS-8700&150.09740 &2.26848 &B&...\\    
COSMOS-10599&150.09535 &2.28725 &B&[O III]\\
COSMOS-11530&150.11931 &2.29688 &B&...\\  
COSMOS-12102&150.09728 &2.30252 &B&[O III], H$\alpha$\\
COSMOS-13184&150.12424 &2.31367 &B&[O III]\\    
COSMOS-14249&150.11011 &2.32459 &B&...\\
COSMOS-14435&150.16232&2.32602&D&...\\
COSMOS-15091&150.15955&2.33330&D&[O III]\\
COSMOS-16152&150.18762 &2.34469&A&...\\
COSMOS-16286&150.17699&2.34539&D&[O III]\tablenotemark{b}\\
COSMOS-16566&150.17067&2.34830&D&[O III], H$\alpha$\\
COSMOS-17118&150.15114&2.35410&D&[O III]\\
COSMOS-17162&150.15134 &2.35482&A&...\\
COSMOS-17295&150.18318&2.35537&D&...\\
COSMOS-17539&150.12814 &2.35810&A&...\\
COSMOS-17839&150.15677 &2.36080&A&[O III]\\
COSMOS-18299&150.17098 &2.36536&A&...\\
COSMOS-18358&150.16719 &2.36689&A&[O III]\\
COSMOS-18582&150.13281 &2.36878&A&...\\
COSMOS-18777&150.18628 &2.37054&A&...\\
COSMOS-19049&150.13886 &2.37340&A&[O III]\tablenotemark{b}, H$\alpha$\\
COSMOS-19077&150.18309 &2.37295&A&[O III]\\
COSMOS-20589&150.18056&2.38822&D&... \enddata
\tablecomments{All IDs refer to the CANDELS catalog for that particular field.  Mask A was observed on 21 April 2012, mask B on 22 April 2012, mask C on 10 and 11 October 2012, and mask D on 12 March 2013, all at the LBT; \textit{GOODS-S-7892} was observed on 15 December 2012, \textit{GOODS-S-43693} on 1 October 2012, \textit{GOODS-S-43928} on 15 October 2012, \textit{UDS-6195} and \textit{UDS-6377} on 27 August 2012, and \textit{UDS-12539} on 2 and 27 August 2012 (120 minutes in the NIR), all at the VLT.  Line detections are at least 1$\sigma$.}
\tablenotetext{a}{Due to technical problems during the first night of observations, the total exposure time used for these line extractions is only 3600s in $H$.}
\tablenotetext{b}{Only the [O III]$\lambda$5007 component.}
\end{deluxetable}
\subsubsection{LUCI1 Data Reduction}
We first mask regions of the spectra that are affected by persistence due to the acquisition and alignment exposures.  This effect is reduced with each readout, so we only mask the regions if they are 2$\sigma$ higher than the background level.  We then create flat-field images from lamp-illuminated exposures and remove cosmic rays using a median-stacking technique.  The most important cosmetic step is the removal of bad pixels, which are identified in the lamp-illuminated exposures, and the hot pixels in the spectra, which are identified in dark exposures.  Additionally, for an as-yet unknown reason, the first exposure of every series has small ``halos'' around the hot pixels.  As such, we remove slightly larger regions around these hot pixels in the first exposure of every series.  Our wavelength calibration is done using the OH sky lines, with a code based on \texttt{XIDL} routines\footnote{\url{http://www.ucolick.org/
~xavier/IDL/}}.  We also use \texttt{XIDL} for the final sky subtraction, which uses a spline-fitting algorithm to measure and remove the lines.  To maximize S/N, we do not use frame-frame subtraction and instead measure the sky from the individual frames, with the exception being some objects with particularly bad skyline contamination, where frame-frame subtraction better removes the skylines but adds noise to the spectrum.  Dithering the exposures by 3$''$ ensures a decreased dependence on the pixel-to-pixel variations in the detector.

Since our objects of interest have virtually no visible continuum, one-dimensional spectral extraction must be done carefully.  We must visually search for the lines in the stacked reduced spectra.  Since we know the wavelengths of the brightest lines from the grism data, this exercise is straightforward.  We isolate the line region according to a signal-to-noise cut, and then collapse that region in the wavelength direction to create a slice containing the spatial line profile.  This profile is fit with a standard Gaussian function.  The width of the distribution, $\sigma$, and the center, $\mu$, are then used in the full spectral extraction: a Gaussian function with these same $\sigma$ and $\mu$ values is fit for each pixel row in the spatial direction tracing a constant distance from the edge of the (curved) slit with the amplitude as the only free parameter, reflecting the electron counts at that particular wavelength.  All sets of observations were reduced and analyzed separately and the resolution is incorporated into the calculation of the intrinsic line widths to remove systematics in the velocity dispersion measurements.  Flux calibration is done by comparing the integrated counts from the LUCI1 spectrum to the integrated line flux from the 3D-HST grism spectrum when possible.

\subsection{VLT/X-Shooter Spectroscopy}

For an additional six sources, we obtained near-IR and visible spectra using the X-Shooter spectrograph \citep{vernet} on the 8.2 m \textit{ESO Very Large Telescope} (VLT).  We observed five of the objects initially found in \citet{vdw}:  three with previous grism-spectroscopic confirmation in \citet{straughn11} and Weiner et al. (in prep.), and the two remaining candidates with the largest photometrically inferred line fluxes.  A sixth candidate was selected from the \citet{straughn11} sample.  Observations were done in long-slit mode from August to December 2012 with 40 minute integrations using 1$''$/0.9$''$/0.9$''$ (UVB/VIS/NIR) slits and the 100k/1pt/hg/1$\times$2 readout mode.  See Table \ref{tab:obs} for the targets and observing dates.  The proximity of objects \textit{UDS-6377} and \textit{UDS-6195} allowed for them to be observed in the same slit.

Although the X-Shooter spectrograph also observes in the UV-Blue, four of our six objects were not observed during dark time, rendering the data unusable.  The near-IR region of X-Shooter spans the combined $YJHK$ region from 1024$-$2048 nm, while the visible region spans 559.5$-$1024 nm.  Reduction of the X-Shooter data is performed using version 2.0.0 of the ESO XSHOOTER pipeline\footnote{\url{http://www.eso.org/sci/software/pipelines/xshooter/xsh-pipe-recipes.html}}, which provides merged, 2D near-IR and visible spectra.  Extraction is performed in a similar manner to the LUCI1 data.

\section{Dynamical and Stellar Masses}
In order to confirm our hypothesis that these systems represent starbursting dwarf galaxies, we must confirm their low stellar masses, low metallicities, and high star formation rates.  Stellar masses can be constrained through SED fits to broadband photometry, and metallicities can be constrained by observing emission-line ratios, such as [O III]/H$\beta$.  Any star formation rate is contingent on the nature of the emission lines, since AGN can also produce very high excitations.
\subsection{Methods and Results}

\label{sec:lines}

\begin{deluxetable*}{lccccccccc}
\tabletypesize{\scriptsize}
\tablecaption{Sample of Emission Line Galaxies\label{tab:lines}}
\tablewidth{0pt}
\tablehead{ \colhead{ID} & \colhead{m$_{F140W}$} & \colhead{r$_{eff}$\tablenotemark{a}} & \colhead{f$_{H\alpha}$} &\colhead{EW$_{H\alpha}$}&\colhead{f$_{[O III]}$}&\colhead{EW$_{[O III],5007}$}&\colhead{z$_{spec}$}&\colhead{$\sigma_{H\alpha}$}&\colhead{$\sigma_{[O III]}$} \\
    & (AB) & (kpc) & &(\AA) &&(\AA)& &($\kms$) &($\kms$)}
\startdata
 GOODS-S-33131 &  23.66$\pm$0.08&  0.68$\pm$0.62& ... & ... &  7.36$\pm$2.99 &  693$\pm$47&  1.687 & 48.7$\pm$4.3 &   52.3$\pm$5.7 \\
 GOODS-S-43693 &  24.36$\pm$0.12&  0.35$\pm$0.06& ... & ... &  16.9$\pm$0.88 &  861$\pm$66&  1.738 & 54.4$\pm$4.5 &... \\
 GOODS-S-43928 &  24.59$\pm$0.15&  1.9$\pm$0.48& 4.5$\pm$1.2\tablenotemark{b} & 199\tablenotemark{b} &  3.7$\pm$1.6\tablenotemark{b} &  176\tablenotemark{b} &  1.472 &...&  31.4$\pm$8.2  \\
 
 UDS-6195 &  24.26$\pm$0.13&  1.4$\pm$0.42& ... & ... &  ... &  701$\pm$95\tablenotemark{c}&  1.687 & 69.9$\pm$4.9 &  54.7$\pm$6.1 \\
 UDS-6377 &  24.53$\pm$0.17&  0.67$\pm$0.04& ... & ... &  ... &  731$\pm$86\tablenotemark{c}&  1.664 & 54.5$\pm$4.5 &  48.2$\pm$5.9 \\
  UDS-7665 &  25.40$\pm$0.14&  0.51$\pm$0.08& ... & ... &  13.7$\pm$2.76 &  803$\pm$162&  2.298 & ...&  57.8$\pm$9.7 \\
   UDS-10138 &  23.77$\pm$0.03&  0.75$\pm$0.08&  ... &  ...&  10.9$\pm$0.57 &  322$\pm$17 &  2.151 & ...&  80.9$\pm$10.0 \\
    UDS-12435 &  23.42$\pm$0.03&  1.0$\pm$0.06&  ...&  ... &  12.2$\pm$0.94 &  263$\pm$31 &  1.611 &  65.2$\pm$11.3 & ...\\
 UDS-12539 &  23.39$\pm$0.06&  1.3$\pm$0.07& ... & ... &  35.5$\pm$2.73 &  713$\pm$42&  1.621 & 81.3$\pm$4.3 &  71.1$\pm$5.7\\
  UDS-19167 &  23.99$\pm$0.04&  1.1$\pm$0.20 &  ...&  ...&  21.5$\pm$2.82&  723$\pm$95 &  2.185 & ...&  54.2$\pm$9.4 \\
  UDS-24154 &  23.78$\pm$0.04 &  1.8$\pm$0.16 &  ...&  ... &  21.9$\pm$3.09 &  503$\pm$34 &  2.297 & 72.5$\pm$13.1 &  61.0$\pm$10.8 \\
COSMOS-10599 &  24.47$\pm$0.26  &  0.67$\pm$0.16&...& ...&  11.8$\pm$0.89 &  714$\pm$85 &  2.220 & ...&  30.9$\pm$9.0 \\
COSMOS-12102 &  22.82$\pm$0.06 & 1.6$\pm$0.08 &  19.9$\pm$1.0 &  360$\pm$18 &  49.4$\pm$2.28 &  630$\pm$29 &  1.463 &  230.8$\pm$14.7 &  241.3$\pm$12.7 \\
COSMOS-13184 &  23.91$\pm$0.16 &  0.53$\pm$0.10 &... & ... &  11.5$\pm$2.94 &  598$\pm$189 &  2.199 & ...&  40.3$\pm$8.9 \\
COSMOS-15091 &  25.46$\pm$0.13& 0.75$\pm$0.11& ... & ... &  5.99$\pm$3.61 &  628$\pm$152&  1.583 & ...&  38.2$\pm$10.0 \\
COSMOS-16286 &  24.64$\pm$0.11& 1.1$\pm$0.13& 0.39$\pm$3.32 & 41$\pm$345 &  8.76$\pm$3.46 &  888$\pm$351&  1.444 & ...&  46.7$\pm$14.4 \\
COSMOS-16566 & 24.60$\pm$0.09&  1.4$\pm$0.09& 4.22$\pm$3.26 & 560$\pm$432 &  7.28$\pm$3.58 &  388$\pm$191&  1.437 & 25.5$\pm$14.0 &  32.8$\pm$8.4 \\
COSMOS-17118 & 24.16$\pm$0.13 & 4.8$\pm$0.33& ... & ... &  12.3$\pm$4.22 &  493$\pm$169&  1.656 & ...&  46.5$\pm$8.8 \\
COSMOS-17839& 24.36$\pm$0.24 &  0.99$\pm$0.06 & 3.27$\pm$3.20 &  325$\pm$230 &  16.3$\pm$3.58 &  1126$\pm$247 &  1.412 & ...&  43.3$\pm$8.9 \\
COSMOS-18358 &  22.84$\pm$0.04 &  1.6$\pm$0.02& ... & ... &  30.2$\pm$0.97 &  316$\pm$26 &  1.645 & ...&  55.9$\pm$9.0 \\
 COSMOS-19049 &  22.93$\pm$0.05 &  2.3$\pm$0.06 &  16.4$\pm$1.2 &  368$\pm$28 &  20.9$\pm$2.20 &  330$\pm$35 &  1.370 &  81.9$\pm$50.2 &  122.0$\pm$11.0\\
 COSMOS-19077 &  23.69$\pm$0.12 &  1.6$\pm$0.05 & ...& ... &  20.5$\pm$0.77 &  536$\pm$20 &  1.649 & ...&  47.7$\pm$9.5\enddata
\tablecomments{All fluxes are given in units of 10$^{-17}$ erg s$^{-1}$ cm$^{-2}$.  Equivalent widths are quoted in the rest-frame.  A description of the size measurements is given in Section \ref{sec:lines}.}
\tablenotetext{a}{\citet{vdw12}}
\tablenotetext{b}{\citet{straughn11}}
\tablenotetext{c}{\citet{vdw}}
\end{deluxetable*}

In our near-IR spectra, the most prominent lines seen are [O III]$\lambda$5007 and H$\alpha$, along with their associated complexes ([O III]$\lambda$4959+H$\beta$, and [N II]$\lambda\lambda$6548,6584, respectively).  With the exception of \textit{COSMOS-12102}, the emission lines can be well-fit by a Gaussian function, as described in \citet{maseda}.  \textit{COSMOS-12102} has the broadest emission lines of the sample.  In addition to their broadness, they display some degree of asymmetry and are thus not well-fit by Gaussian functions.  
  
The skewness could be caused by several processes, such as the presence of strong outflows.  Such interpretations are beyond the scope of this paper and demand additional observations.

The best-fitting redshifts, velocity dispersions, and line ratios are given in Table \ref{tab:lines}.

As described in \citet{maseda}, velocity dispersions of the strong emission lines can be used to estimate the dynamical masses, assuming the line width comes entirely from gravitational motion in a virialized system such that
\begin{equation}
 M_{dyn} = 3\frac{r_{\rm{eff}}\sigma^2}{G}.
\label{eqn:dyn}
\end{equation}
Here, $\sigma$ is the observed line width from our NIR spectrum and $r_{eff}$ is the effective radius of the galaxy from the public CANDELS catalog released in \citet{vdw12}.  Typical objects are 1.0$\pm$0.1 kpc in both the $J_{F125W}$- and $H_{F160W}$-bands.

These dynamical masses are listed in Table \ref{tab:dp}, ranging from $10^{8.39}$ to $10^{10.6} ~\msol$, with a median mass of $10^{9.13}~\msol$.  The uncertainty in the dynamical mass estimate comes primarily from the systematic uncertainty in the proportionality constant of 3, which relates the intrinsic velocity $v$ to the observed velocity dispersion $\sigma$, which we assume to be the same factor of 33\% as \citet{rix}, since in most cases our observed line widths and physical sizes are well-constrained.  Further details can be found in \citet{maseda}.  \citet{amorin12}, in a study of ``green peas,'' observe multiple star-forming regions and gas flows which is seen as asymmetries and broad, low-intensity wings.  While we do not see such clear evidence for outflows via asymmetric line profiles (\textit{COSMOS-12102} excepted) or broad wings, we can not currently rule-out contributions of non-gravitational motions to the observed line widths since we do not resolve the continuum and can only observe the bright, central line regions.  Any such contributions would mean that the intrinsic dispersion is lower and that our dynamical mass estimates are upper-limits.

\label{sec:sed}
Multi-band photometry is obtained from 3D-HST \citep{skelton}, covering 0.3-24 $\mu$m for the GOODS-S (23 bands), UDS (17 bands), and COSMOS (31 bands) fields.  Visual inspection of the Spitzer-IRAC frames show contamination from nearby objects in \textit{UDS-7665}, \textit{COSMOS-13184}, and \textit{COSMOS-15091}, so their SEDs do not include the contaminated points.  For the same reason, we do not include any the data from the 5.8 and 8$~\mu$m IRAC channels for this sample, even though it is available as part of the publicly-released catalogs.  Only three objects have detections at 24 $\mu$m with Spitzer-MIPS (discussed in Section \ref{sec:starburstagn}), and an upper limit of 10 $\mu$Jy is adopted for the remainder of the sample.

We fit the broadband spectral energy distributions of our galaxies in the same manner as \citet{maseda} using a custom version of the \texttt{MAGPHYS} code\footnote{\url{http://www.iap.fr/magphys}} \citep{magphys}, which computes the emission by stellar populations and the attenuation by dust in a two-component ISM, and includes nebular line emission computed self-consistently using the \citet{pacifici12} model (\citeauthor{cl} \citeyear{cl}; C. Pacifici et al in prep).  The broad-band fluxes computed with this model include the contamination by emission lines, so they can be directly and robustly compared with the observed fluxes that we know are likely emission-line contaminated for these galaxies \cite[at these and higher redshifts, it has been shown that an improper treatment of nebular contamination to broadband magnitudes results in an overestimate of stellar masses and hence an underestimate in sSFRs, e.g.][]{atek,curtislake,stark}.  \texttt{MAGPHYS} compares the input photometry to an extensive library of SED templates spanning a wide range in parameters such as star formation history, metallicity, age, and dust optical depth using a Bayesian method.  As such, all results quoted are the medians of the posterior probability distributions 
for each parameter, with 
uncertainties corresponding to the 16$^{th}$ and 84$^{th}$ percentiles for the distribution.  In cases where the output probabilities are not well constrained, typically due to the (potentially sytematic) uncertainties in the photometry, we adopt an uncertainty of 0.3 dex in the relevant parameter: a formal error of 0 is indicative of the models not fitting the data well.  0.3 dex is the typical uncertainty in determining stellar masses from fits to broadband photometry \citep{conroy}.  An example probability distribution for some of the various parameters is shown in the Appendix, Figure \ref{fig:pdf}. Results of the SED fitting are given in Table \ref{tab:dp}, showing high sSFRs, low stellar masses, low metallicities, young ages, and low dust extinction.  Since the NIR sizes represent the restframe optical and hence the stellar continuum at these redshifts, we are probing the same physical region in both mass estimates.

The median $\tau_V$ ($V$-band optical depth seen by young stars in the birth clouds) is 0.2, consistent with the very blue observed SEDs.  Even with the lack of infrared data to directly probe the dust content of these systems, we can place a limit on the dust mass based on the total dust attenuation and luminosity inferred from the SED fits and a prior on the dust temperature as in \cite{dacunha}.  Resulting limits are $\lesssim 10^7~\msol$ and hence negligible compared to the stellar masses.

As mentioned before, the critical piece of additional information that we include to perform these fits is the line fluxes.  We see a median SFR of $\sim$9 $\msol~yr^{-1}$ which, combined with the low stellar masses, justifies our emission line criteria for the selection of starbursts. By separating the emission lines from the stellar continuum light, we are better able to trace the gas-phase metallicities.  This results in metallicity estimates consistent with direct probes of the oxygen abundance using emission-line ratios (see Section \ref{sec:metals}).  In addition, it allows for better estimates of the extinction in the HII regions, which produce the aforementioned $\tau_V$ values.

Our model library of stellar population SEDs contains a broad range of
complex star formation histories (SFHs), including bursts on top of
extended SFHs with a variety of evolutionary trends (rising, falling,
and constant).  Despite these efforts, which far exceed the still
common use of exclusively exponentially-declining SFHs, systematic
uncertainties remain.  In particular for galaxies with significant
star formation activity in the past $\sim 50$~Myr, as is the case
here, red supergiants with individual luminosities of $\sim
10^5~L_{\odot}$ can easily outshine more massive populations of stars
with any age $>50$~Myr, especially in the near-infrared.  Prior
knowledge of the SFH is needed to address this issue, producing a
degree of circularity in the problem of stellar mass determinations.
Keeping this in mind, we proceed and note where necessary that for
galaxies with estimated ages $\lesssim 50$~Myr the mass (and age)
estimates must be lower limits, as seen in tails to higher masses and 
mass-weighted ages, e.g. Figure \ref{fig:pdf}.

\begin{figure}
 \includegraphics[width=.475\textwidth]{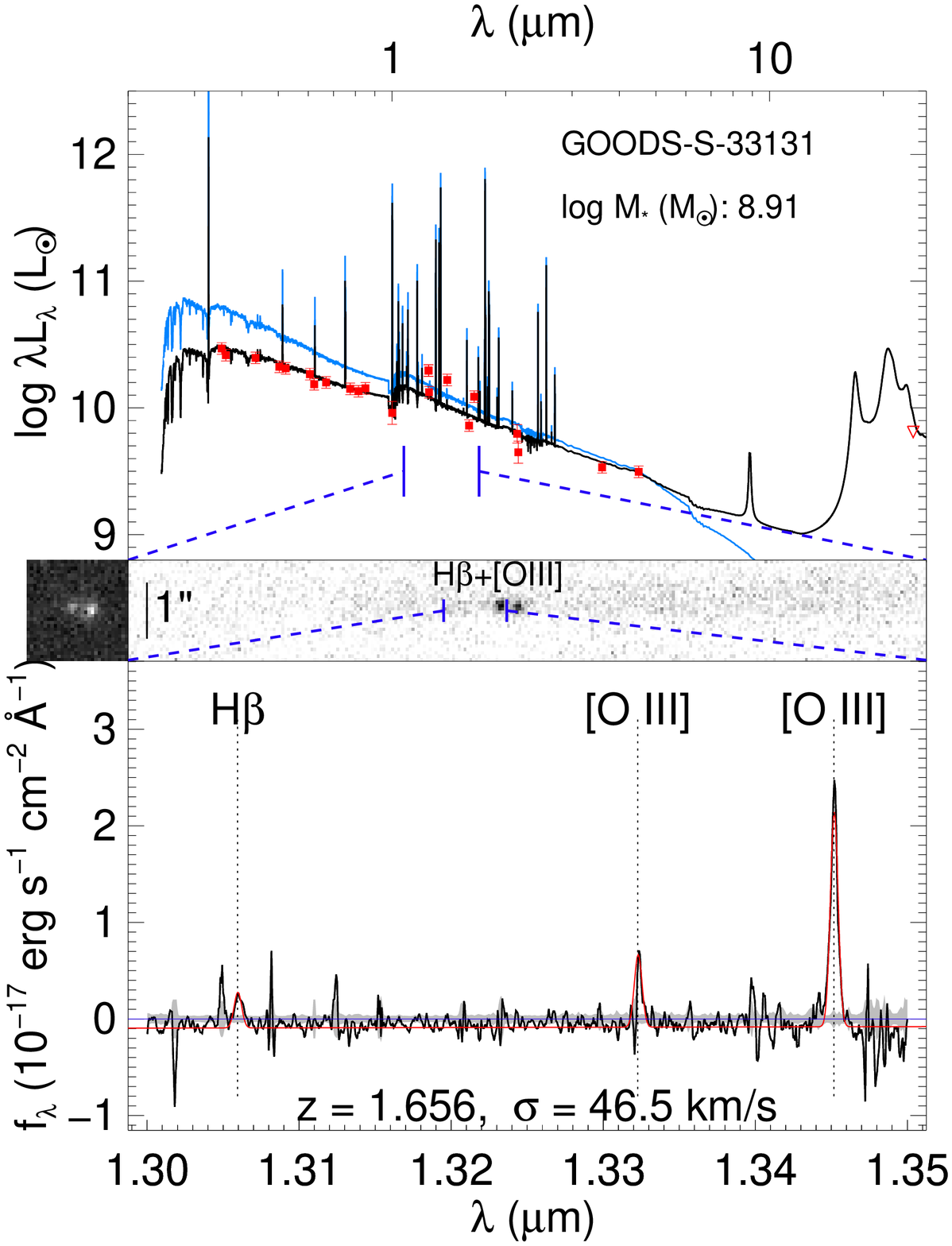}

\caption{Example of a broadband SED including best-fitting model (top), WFC3 grism image/spectrum (middle), and X-Shooter spectrum (bottom) for an object in our sample.  The NIR spectra for the remaining objects can be seen in the Figures 2a for [O III] and 2b for H$\alpha$; the SEDs for the remaining objects can be seen in the Appendix, Figure \ref{fig:seds}.  SED fits are performed as described in Section \ref{sec:sed}, with red points denoting the measured photometry (open points are upper limits), the blue curve denoting the non-attenuated SED, and the black curve denoting the observed SED including dust attenuation.  The direct F140W image is shown on the left and the dispersed G141 grism image is shown to the right, with important spectral lines labeled and contamination subtracted as described in Momcheva et al. (in prep.).  The X-Shooter spectrum is smoothed by 3 pixels and is flux-calibrated according to the grism line flux.  The shaded gray area represents the +/$-$ 1$\sigma$ flux uncertainties and the red curve shows the best-
fitting model of 
the emission lines.}
\label{fig:caspec}

\end{figure}

Extracted near-IR spectra and SED fitting results for \textit{GOODS-S-33131} are shown in Figure \ref{fig:caspec}; all remaining objects in our sample are shown in Figures 2a, 2b, and \ref{fig:seds}.  Telluric corrections are applied as needed.

\begin{deluxetable*}{lcccccc}
\tabletypesize{\scriptsize}
\tablecaption{Derived Parameters\label{tab:dp}}
\tablewidth{0pt}
 \tablehead{ \colhead{ID} & \colhead{log $M_{dyn}$} & \colhead{log $M_{\star}$} & \colhead{log Age} & \colhead{Z} & \colhead{log SFR} & \colhead{$\tau_{V}$} \\
 &($\msol$)&($\msol$)&(yr)&($\zsol)$&($\msol~$yr$^{-1}$)&}
\startdata
GOODS-S-33131 & 9.05$\pm$0.30&8.91$_{-0.075}^{+0.050}$&7.78$_{-0.065}^{+0.190}$&0.321$_{-0.132}^{+0.180}$&0.892$_{-0.080}^{+0.030}$&0.242$_{-0.085}^{+0.030}$\\
GOODS-S-43693 & 8.86$\pm$0.31&8.28$_{-0.050}^{+0.195}$&7.50$_{-0.045}^{+0.240}$&0.147$_{-0.098}^{+0.160}$&0.557$_{-0.065}^{+0.090}$&0.137$_{-0.090}^{+0.050}$\\
GOODS-S-43928 & 9.12$\pm$0.38&8.83$_{-0.065}^{+0.070}$&8.39$_{-0.260}^{+0.175}$&0.169$_{-0.120}^{+0.272}$&0.262$_{-0.080}^{+0.190}$&0.327$_{-0.145}^{+0.505}$\\
UDS-6195 & 9.47$\pm$0.33&7.60$_{-0.300}^{+0.320}$&6.85$_{-0.300}^{+0.385}$&0.299$_{-0.130}^{+0.300}$&0.447$_{-0.065}^{+0.300}$&0.027$_{-0.300}^{+0.085}$\\
UDS-6377 & 9.04$\pm$0.31&8.32$_{-0.300}^{+0.140}$&6.96$_{-0.300}^{+0.735}$&0.081$_{-0.022}^{+0.096}$&1.08$_{-0.580}^{+0.300}$&1.12$_{-0.835}^{+0.025}$\\
UDS-7665& 9.07$\pm$0.33& 8.52$_{-0.300}^{+0.300}$& 7.04$_{-0.300}^{+0.300}$& 0.155$_{-0.300}^{+0.300}$& 1.21$_{-0.300}^{+0.300}$& 0.862$_{-0.300}^{+0.300}$\\
UDS-10138 & 9.53$\pm$0.31&8.43$_{-0.300}^{+0.270}$&7.10$_{-0.300}^{+0.430}$&0.177$_{-0.300}^{+0.010}$& 1.05$_{-0.135}^{+0.300}$&0.112$_{-0.040}^{+0.300}$\\
UDS-12435 & 9.47$\pm$0.33&9.42$_{-0.055}^{+0.045}$&8.60$_{-0.130}^{+0.050}$&0.189$_{-0.138}^{+0.392}$& 0.677$_{-0.030}^{+0.070}$&0.082$_{-0.040}^{+0.085}$\\
UDS-12539& 9.66$\pm$0.30&8.67$_{-0.300}^{+0.300}$&7.05$_{-0.300}^{+0.300}$&0.581$_{-0.300}^{+0.300}$&1.29$_{-0.300}^{+0.300}$&0.667$_{-0.300}^{+0.300}$\\
UDS-19167& 9.35$\pm$0.34& 8.96$_{-0.300}^{+0.300}$ & 8.58$_{-0.300}^{+0.300}$& 0.085$_{-0.300}^{+0.300}$& 1.02$_{-0.300}^{+0.300}$& 0.192$_{-0.300}^{+0.300}$\\
UDS-24154 & 9.67$\pm$0.33&9.10$_{-0.300}^{+0.100}$&7.49$_{-0.300}^{+0.165}$&0.531$_{-0.030}^{+0.300}$& 1.38$_{-0.050}^{+0.300}$&0.462$_{-0.005}^{+0.105}$\\
COSMOS-10599 & 8.65$\pm$0.40&8.72$_{-0.300}^{+0.240}$& 7.49$_{-0.300}^{+0.245}$ & 0.531$_{-0.362}^{+0.300}$& 1.00$_{-0.020}^{+0.015}$&0.462$_{-0.300}^{+0.080}$\\
COSMOS-12102& 10.8$\pm$0.30&9.57$_{-0.300}^{+0.300}$& 8.40$_{-0.300}^{+0.300}$& 0.155$_{-0.300}^{+0.300}$& 1.63$_{-0.300}^{+0.300}$&1.77$_{-0.300}^{+0.300}$\\
COSMOS-13184 & 8.78$\pm$0.36&9.00$_{-0.220}^{+0.070}$& 7.28$_{-0.095}^{+0.365}$ & 0.087$_{-0.040}^{+0.068}$& 1.29$_{-0.120}^{+0.230}$&0.407$_{-0.110}^{+0.245}$\\
COSMOS-15091& 8.88$\pm$0.37&7.92$_{-0.120}^{+0.280}$& 7.61$_{-0.105}^{+0.965}$ & 0.169$_{-0.084}^{+0.042}$& 0.062$_{-0.020}^{+0.200}$&0.077$_{-0.300}^{+0.115}$\\
COSMOS-16286 & 9.22$\pm$0.40&8.44$_{-0.125}^{+0.160}$& 7.92$_{-0.650}^{+0.525}$ & 0.171$_{-0.054}^{+0.044}$& 0.317$_{-0.325~}^{+0.440~}$&1.05$_{-0.850}^{+0.500}$\\
COSMOS-16566 & 9.02$\pm$0.37&8.60$_{-0.045}^{+0.040}$& 8.29$_{-0.110}^{+0.065}$ & 0.137$_{-0.074}^{+0.324}$& 0.147$_{-0.045}^{+0.035}$&0.037$_{-0.020}^{+0.050}$\\
COSMOS-17118& 9.86$\pm$0.33&8.47$_{-0.315}^{+0.210}$& 7.66$_{-0.425}^{+0.920}$ & 0.155$_{-0.070}^{+0.136}$& 0.752$_{-0.085}^{+0.300}$&0.192$_{-0.010}^{+0.105}$\\
COSMOS-17839 & 9.11$\pm$0.34&8.19$_{-0.190}^{+0.135}$&7.43$_{-0.195}^{+0.335}$& 0.091$_{-0.018}^{+0.300}$& 0.392$_{-0.035}^{+0.105}$&0.037$_{-0.300}^{+0.005}$\\
COSMOS-18358& 9.54$\pm$0.32& 9.43$_{-0.300}^{+0.300}$& 8.12$_{-0.300}^{+0.300}$ & 0.629$_{-0.300}^{+0.300}$ & 1.11$_{-0.300}^{+0.300}$&0.037$_{-0.300}^{+0.300}$\\
COSMOS-19049 & 10.4$\pm$0.30&9.85$_{-0.300}^{+0.168}$ &8.36$_{-0.300}^{+0.165}$& 0.081$_{-0.300}^{+0.046}$&1.31$_{-0.300}^{+0.120}$&1.72$_{-0.300}^{+0.200}$\\
COSMOS-19077 & 9.40$\pm$0.34&8.92$_{-0.225}^{+0.300}$& 8.05$_{-0.555}^{+0.300}$ & 0.173$_{-0.300}^{+0.358}$& 0.692$_{-0.300}^{+0.285}$&0.197$_{-0.300}^{+0.265}$\enddata
\tablecomments{Quoted values for $M_{\star}$, Age (mass-weighted), Z, SFR, and $\tau_{V}$ are the medians of the probability distributions from \texttt{MAGPHYS} with associated +/-- 1$\sigma$ values.  Cases where we have an uncertainty of 0 occur when the data cannot be well-explained by the models and not when the models constrain the output parameters well, which manifests itself as a large $\chi^2$ value.  As such, we will adopt an uncertainty of 0.3 dex \cite[the typical uncertainty for stellar masses obtained from fitting broadband photometry,][]{conroy} in those cases to be used in the subsequent analysis. }

\end{deluxetable*}

\begin{figure*}
\label{fig:2a}
\epsscale{.98}
\figurenum{2a}
 \plotone{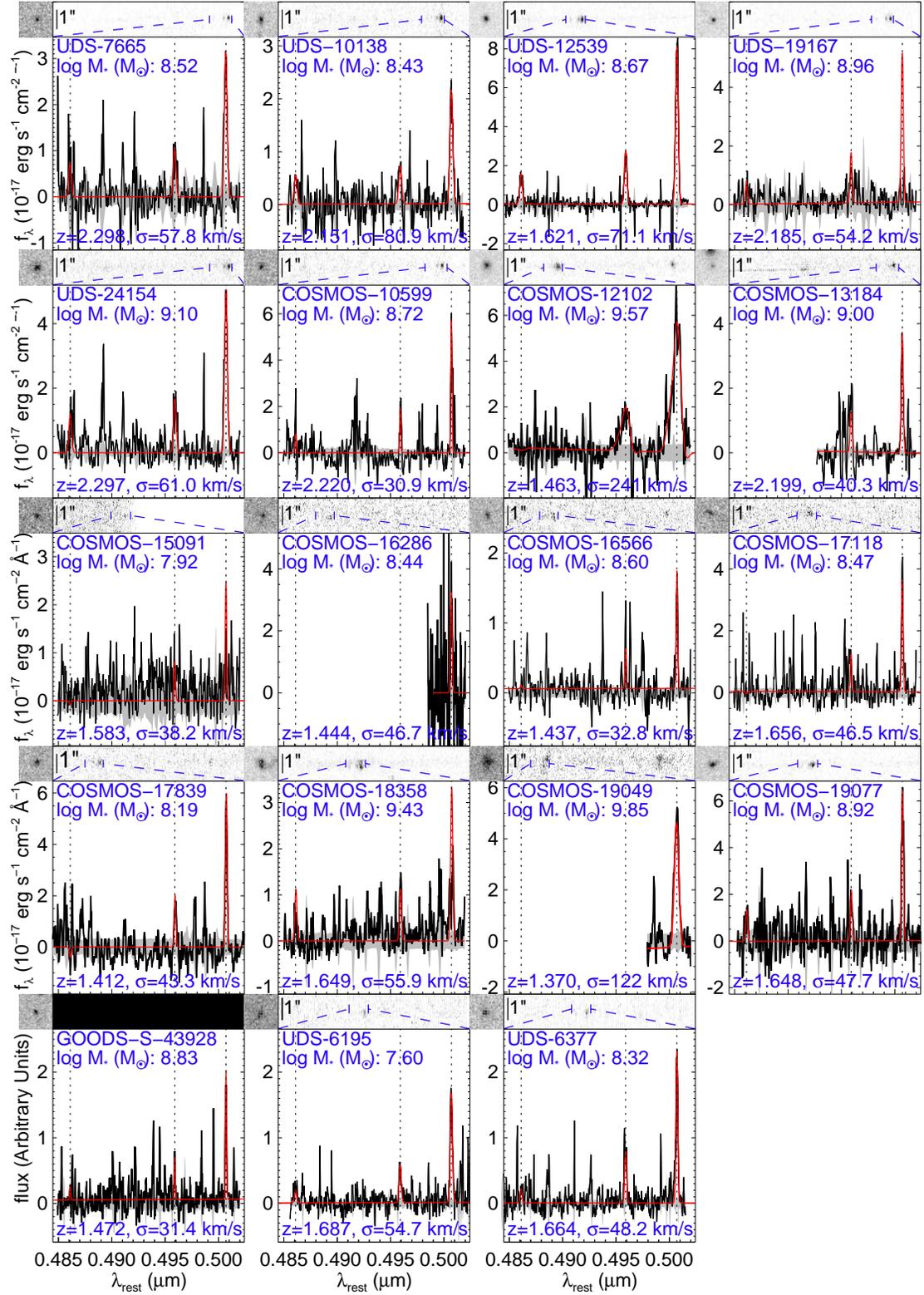}
 \caption{WFC3 G141 grism and LUCI1 or X-Shooter spectrum.  The spectra are smoothed by 3 pixels and are flux-calibrated according to the grism line fluxes.  The shaded gray area represents the +/$-$ 1$\sigma$ flux uncertainties and the red curve shows the best-fitting model of the emission lines.  The dotted lines show the position of the [O III]$\lambda\lambda$5007, 4959 and H$\beta$ emission lines.}
\end{figure*}

 \begin{figure}
 \label{fig:2b}
 \figurenum{2b}
 \epsscale{.98}
  \plotone{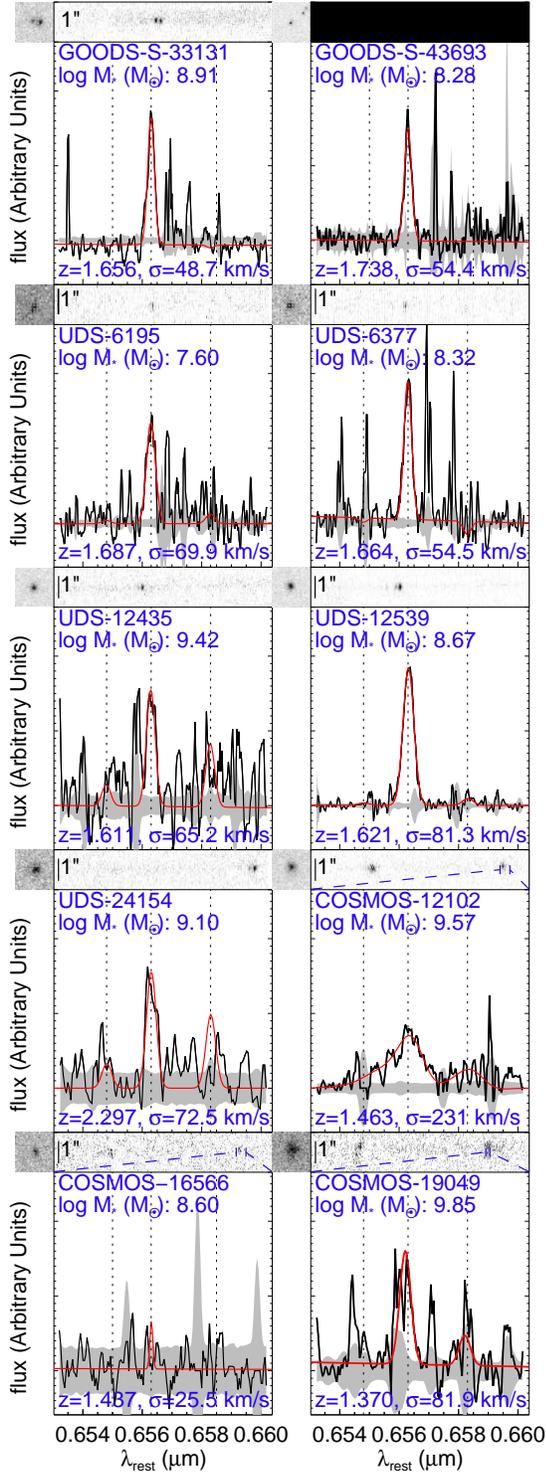}
   \caption{WFC3 G141 grism and LUCI1 or X-Shooter spectrum, same as Figure \ref{fig:2a} but for the detected H$\alpha$ lines.  The positions of the [N II] lines as well as the H$\alpha$ line are denoted by the dotted lines.}
 \end{figure}

\begin{figure}

\includegraphics[width=0.475\textwidth]{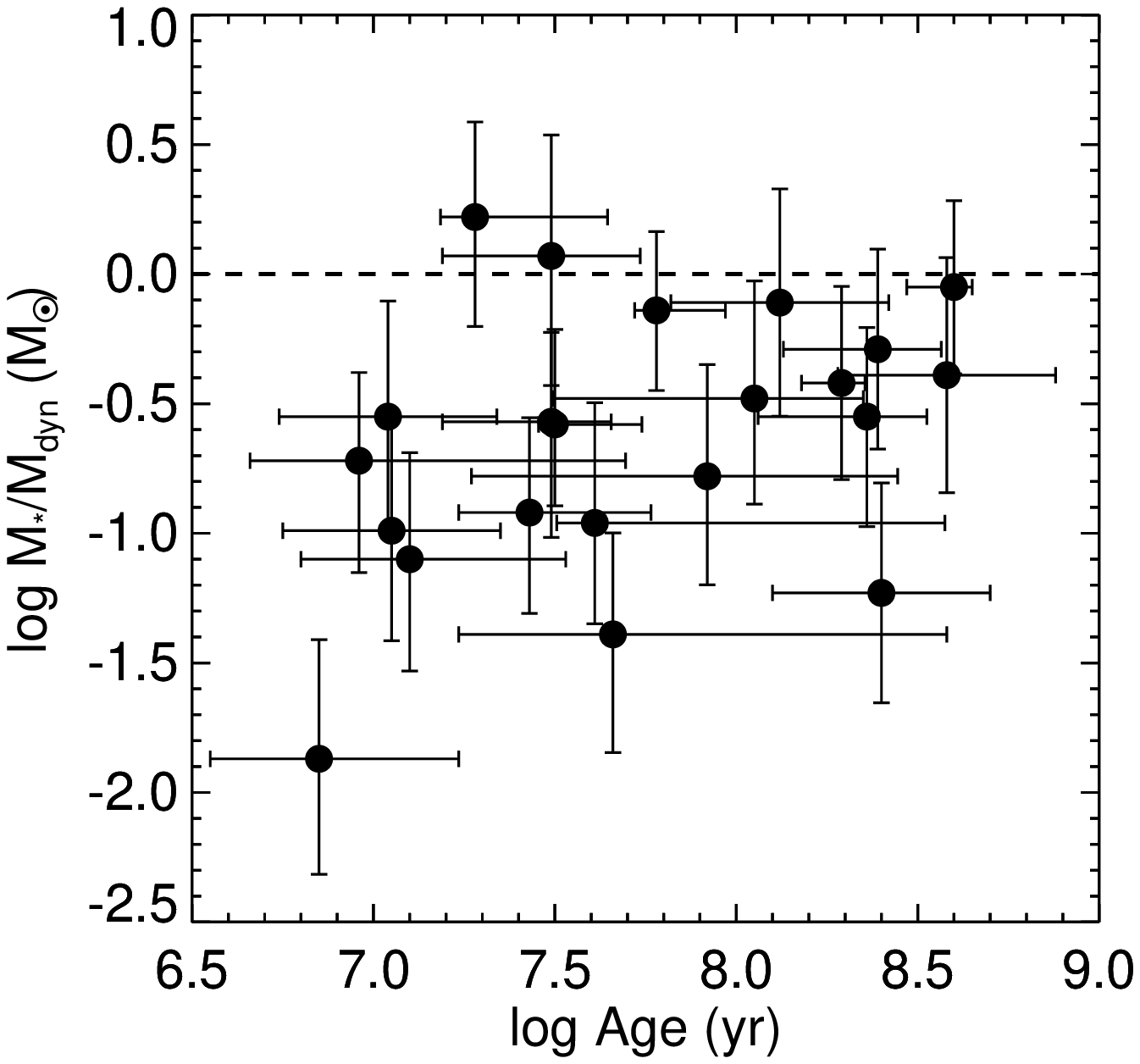}

\caption{Ratio of stellar mass to total dynamical mass versus age (mass-weighted) for the sample, with dynamical masses based on line measurements coming from the ground-based NIR spectra and stellar masses and ages from \texttt{MAGPHYS}.}
\label{fig:dynmass}
\end{figure}

\begin{figure}
 \includegraphics[width=0.475\textwidth]{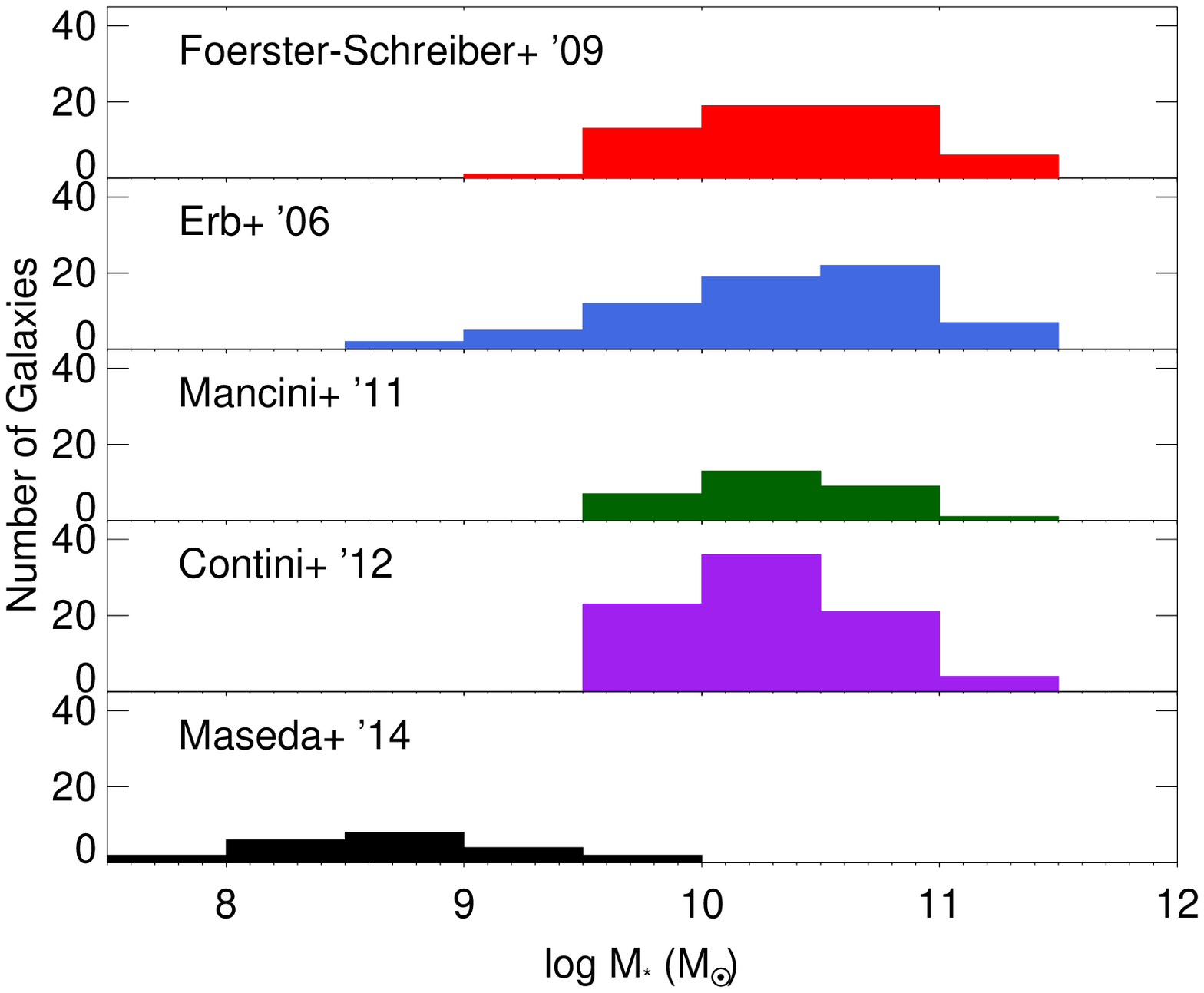}
 \caption{Histograms of stellar mass for our sample compared to the larger H$\alpha$ samples of \citet{fs09}, \citet{erb06}, \citet{mancini11}, and \citet{contini} at $z>1$.  While variations in the stellar templates and the IMF used can alter stellar mass estimates by $\sim$0.3 dex, our sample still lies at considerably lower masses (up to 2 orders of magnitude lower) than the other samples.}
 \label{fig:masshist}
\end{figure}

\begin{figure}

\includegraphics[width=0.475\textwidth]{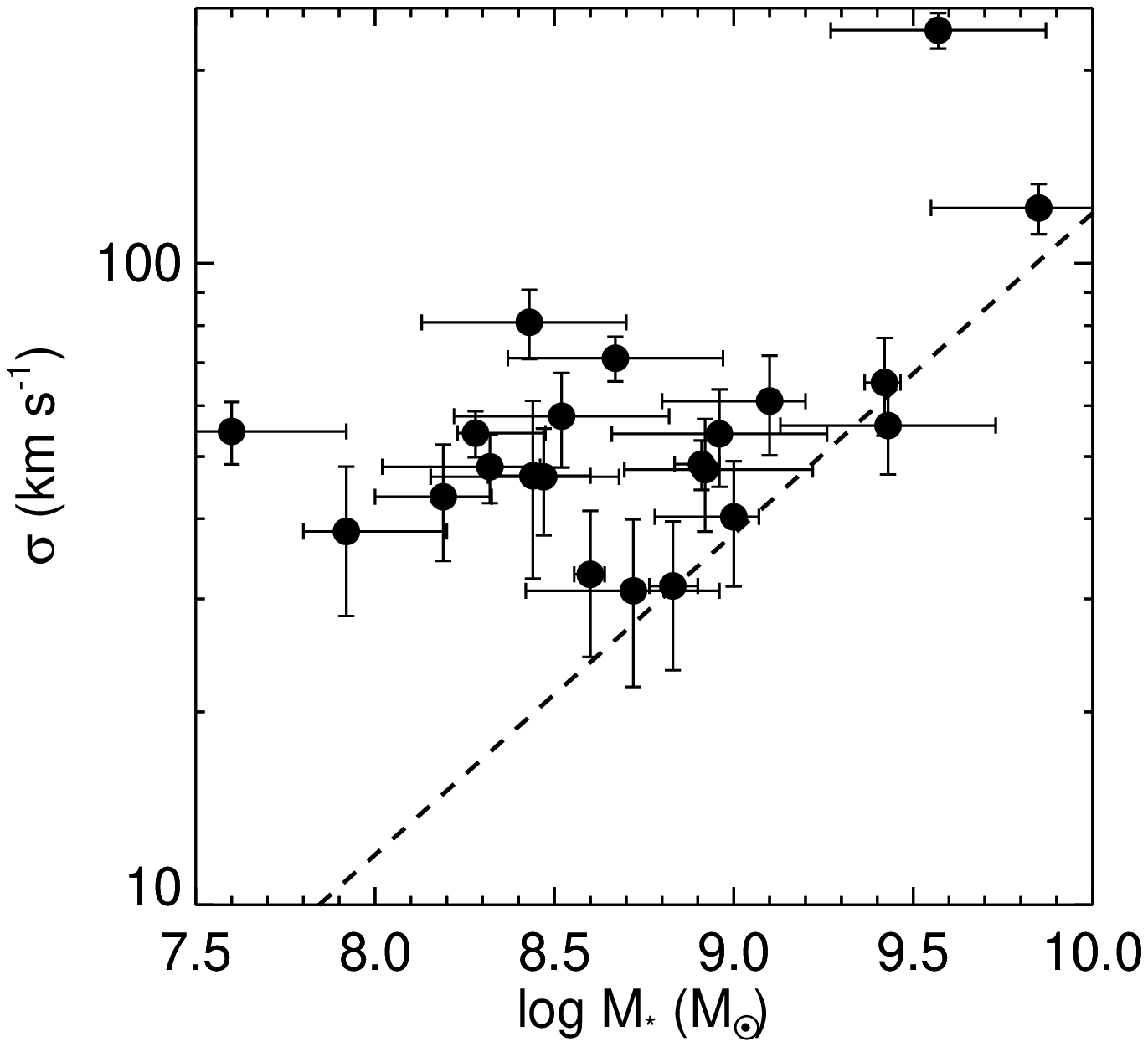}

\caption{Stellar mass versus observed line width for the sample.  There is no clear trend of increasing line width with increasing stellar mass.  The overplotted dashed line corresponds to $M_{\star} \sim M_{virial}$ for a fixed $\reff$ of 1 kpc.}
\label{fig:sigmass}
\end{figure}

\begin{figure}

\includegraphics[width=0.475\textwidth]{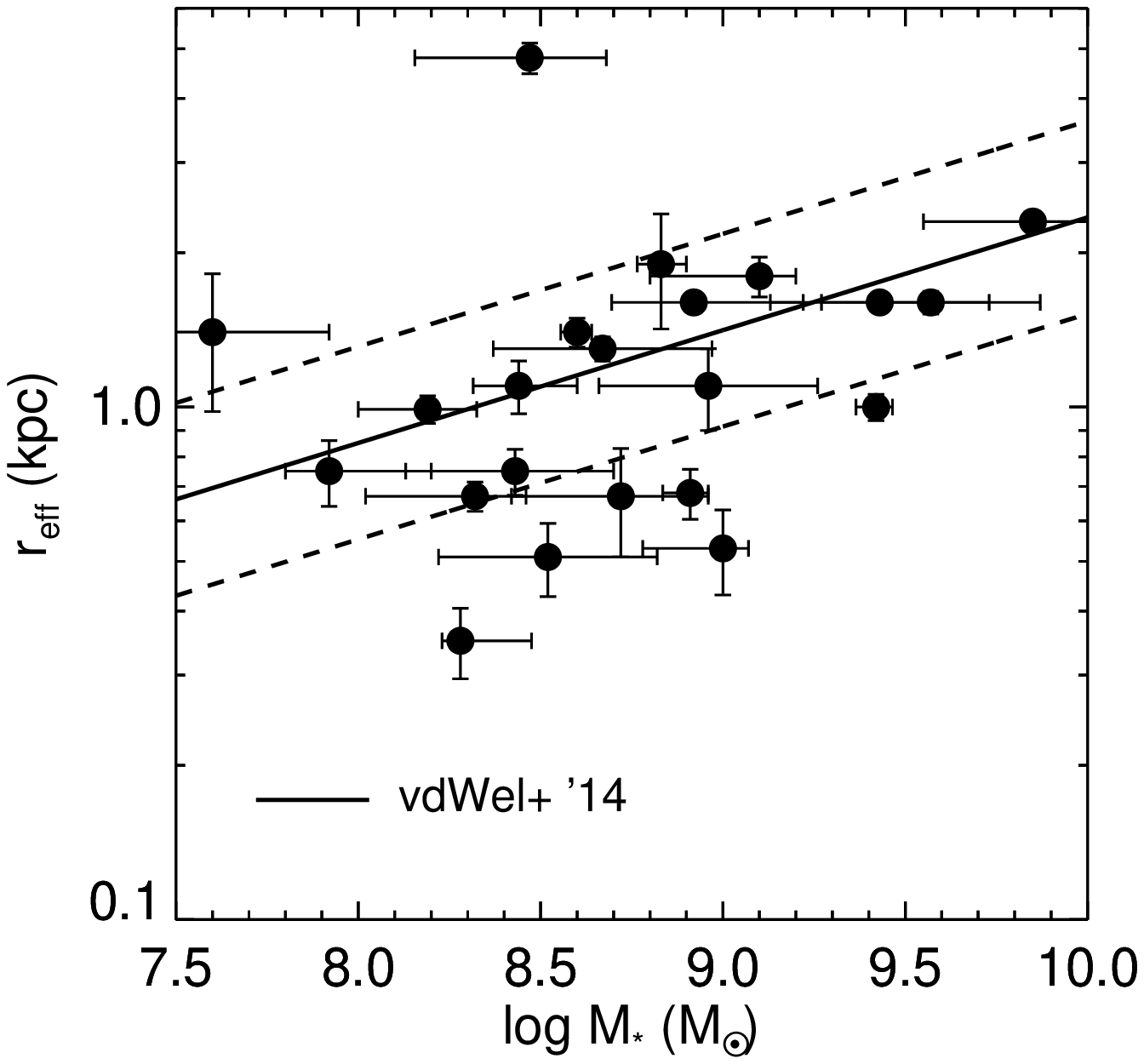}
\caption{Effective radius versus stellar mass for the sample.  The overplotted lines are the low-mass extrapolation (and instrinsic scatter) to the relationship for star-forming galaxies at $2 < z < 2.25$ as found in \citet{vdw14}.  The emission line galaxies in our sample fall on or somewhat below the size-mass relation for normal star-forming galaxies at similar redshifts.}
\label{fig:sizemass}
\end{figure}

As demonstrated in \citet{maseda}, there is a tight, linear relation between the total dynamical and stellar mass estimates.  Figure \ref{fig:dynmass} may show that the stellar-to-dynamical mass ratio correlates with the (mass-weighted) age as well.  This qualitatively agrees with the model in which a non-replenished reservoir of gas is turned into stars, such that the older systems have formed a proportionally larger number of stars.  For further discussion, see Section \ref{sec:discussion}.  To date, these systems represent the lowest dynamical masses ever measured at this epoch, and typically have lower stellar masses than those galaxies in similar star-forming spectroscopic surveys \cite[e.g.][see also Figure \ref{fig:masshist} and Section \ref{sec:comp}]{masters}. Note that the systematic errors in $M_{dyn}$ dominate the size of the error bar.  \textit{COSMOS-12102} shows a broad and asymmetric line profile for which the line width most likely does not trace the underlying gravitational potential.  The lack of a clear trend in the relationship 
between the two main ``observed'' quantities, $M_{\star}$ and $\sigma$, can be seen in Figure \ref{fig:sigmass}, which suggests that any relationship between $M_{\star}$ and $M_{dyn}$ must be driven by a relationship between the size of the galaxy and $M_{\star}$.  Figure \ref{fig:sizemass} shows this relationship, as our sample lies on or below the observed size-mass relation for late-type galaxies at $2 < z < 2.25$ as found in \cite{vdw14} in 3D-HST and CANDELS.

\begin{figure}

 \includegraphics[width=0.475\textwidth]{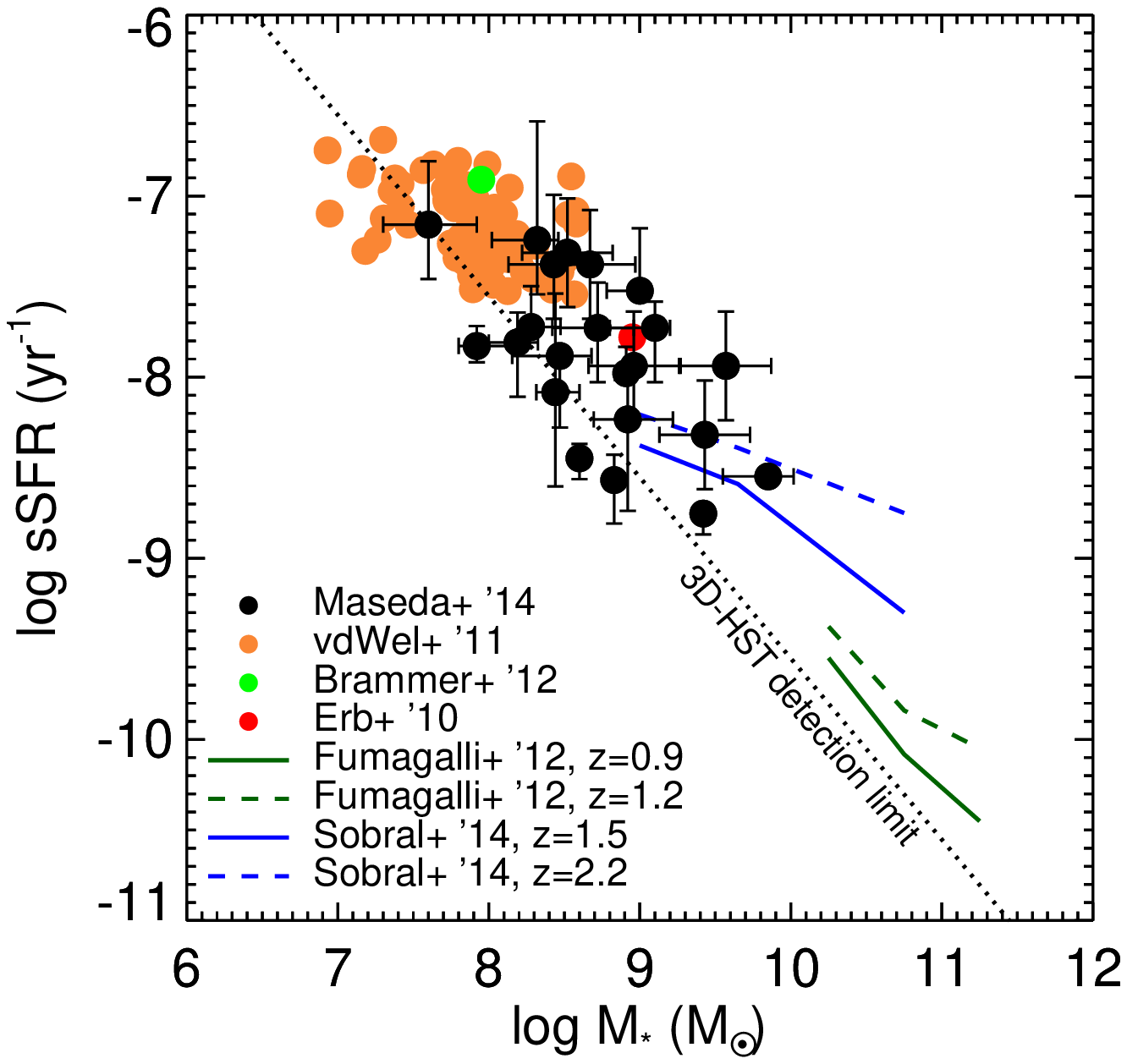}
\caption{sSFR versus stellar mass.  Black points are the results from this study, orange points are the values from \citet{vdw}, the red point is from \citet{erb10}, the green point is from \citet{gb2}, the blue lines are the results for the high/low-z bins of star-forming galaxies in \citet{fumagalli}, and the green lines are the results for the characteristic sSFR (SFR*/M$_{\star}$, see the text for details) for narrow-band selected star-forming galaxies in \citet{sobral}.  The diagonal dotted line represents the nominal detection limit from the 3D-HST survey of 2.8 $\msol$ yr$^{-1}$ at $z \sim 1.5$. The sSFRs are averaged over 10 Myr.}
\label{fig:ssfr}
\end{figure}

The extreme nature of the star-formation in these systems is clearly seen in Figure \ref{fig:ssfr}.  The stellar masses measured here are beginning to probe the same regime as \citet{vdw} and \citet{gb2}, with similar sSFR values.  Note that the sSFR values obtained from \texttt{MAGPHYS} are averaged over the last 10 Myr: given the ephemeral nature of the bursts, the current (s)SFR may not reflect the most vigorous period of star formation in these systems. These are more than an order of magnitude in excess of the sSFR values characteristic of the star-forming population of massive galaxies at similar redshifts, measured in H$\alpha$ from \citet{fumagalli}.

\subsection{Comparison to Other Studies}
\label{sec:comp}
At lower redshifts 0.11 $\leq z \leq$ 0.93, \citet{amorin14a,amorin14b} have isolated a sample of EELGs selected on [O III] flux that show remarkable similarities to our sample with sizes $r_{1/2} \sim$ 1.3 kpc, masses $10^7-10^{10} \msol$, sSFR values $10^{-7}-10^{-9}$ yr$^{-1}$, and metallicity estimates of 0.05 $-$ 0.6 $\zsol$ (some determined ``directly" using the [O III]$\lambda$4363 line as well, see Section \ref{sec:metals}).  Deep HST-ACS $I$-band images reveal that most (80\%) of their EELG sample show non-axisymmetric morphologies indicative of recent mergers or interactions.  Only with samples that are complete over the redshift ranges in question will allow for a direct comparison between the two populations, which are currently only split artificially according to either optical or near-IR observations.  However, caution must be taken in any interpretation: the much higher number density at $z\sim$1.7 from \citet{vdw} than at very low-z from \citet{greenpea} implies that there could be a very different mechanism involved to trigger the bursts.  Connecting the two populations in a qualitative sense is the subject of ongoing work.

Our current observations do not allow us to make strong conclusions about the internal dynamics of individual systems.  Currently, the only opportunities to study the internal dynamics of such small systems at high-redshift is with gravitational lensing: \citet{jones10} note that their $z\sim2-3$ strongly-lensed galaxies would resemble mergers or dispersion-dominated systems without the additional spatial resolution provided by the lensing.  Two objects similar to those presented here are \textit{MACS J2135-0102} ($M_{\star}$=9.8 $\msol$, z=3, $r_{1/2}$=1.75 kpc, SFR=40 $\msol$ yr$^{-1}$, $\sigma_{H\alpha}$=54$\kms$, $V_c$=67$\kms$) from \citet{jones10} and  \textit{SHIZELS-10} ($M_{\star}$=9.4 $\msol$, z=1.45, $r_{1/2}$=2.3 kpc, SFR=10 $\msol$ yr$^{-1}$, $\sigma_{H\alpha}$=64$\kms$, $V_c$=26$\kms$) from \citet{swinbank12}, which both appear to have a (relatively weak) rotational component to their dynamics.  Further discussion of the dynamics of our present sample is deferred to Section \ref{sec:discussion}.

Several other studies have begun to characterize the starforming properties of high-z EELGs using various techniques.  The most obvious comparable study to this work is the WFC3 Infrared Spectroscopic Parallels survey \cite[WISP,][]{wisp}, and specifically the study of \citet{masters}.  As a similar WFC3 grism survey, they are also able to select galaxies based on emission lines instead of photometric techniques, and thus can isolate a sample of high-EW ELGs.  \citet{masters} present a sample of 26 such ELGs with a median restframe [O III] EW of 154 \AA $ $ (our median is 629 \AA).  These galaxies show similarly-low velocity dispersions ($\sim$ 70 $\kms$) and hence also have dynamical masses $\lesssim$ 10$^{10} \msol$.  They derive stellar masses using an assumed M/L ratio and star-formation history in a similar fashion to \citet{vdw}, which have been shown to generally agree with SED-derived stellar masses \citep{maseda}, and have also begun to probe the $M_{\star} \lesssim 10^9 \msol$ regime.  Specific discussion of their metallicity estimates is deferred to Section \ref{sec:metals}.

In addition to WISP, narrowband studies have also begun to uncover the general starforming population of galaxies at $z > 1$, probing stellar masses below 10$^{9.5} \msol$.  \citet{sobral} use data from the HiZELS survey to study H$\alpha$ emitters at redshifts $z = 0.84, 1.47$, and $2.23$.  The size of their survey area ($\sim$ 2 deg$^2$) and the depth of the imaging allows them to isolate large and pure samples of H$\alpha$ emitters down to a restframe H$\alpha$+[NII] EW of 25 \AA, constraining the stellar mass function of star-forming galaxies down to 10$^{9.5} \msol$ at these redshifts.  Indeed, their sample also includes a number of objects with restframe H$\alpha$+[NII] EW values in excess of 300 \AA $ $ and with stellar masses from SED-fitting below 10$^{9} \msol$.  Their results for the characteristic sSFR (i.e. the typical SFR for a galaxy at a given mass and redshift divided by its mass) as a function of mass and redshift is shown in Figure \ref{fig:ssfr}.  While some of our objects can be considered ``normal" at these redshifts according to this determination, a majority of them still have higher sSFRs than expected, albeit typically within 1 dex.  This reinforces the notion that these objects are the high-EW tail of the total distribution of star-forming galaxies at these redshifts and do not comprise a genuinely separate population \citep{vdw}.

\section{Emission-Line Ratios}
\subsection{Starbursts or AGN?}
\label{sec:starburstagn}
So far, the main caveat is that we have assumed that star formation is primarily responsible for the strong line emission.  The most compelling evidence to support this assumption is the relation between stellar mass and dynamical mass, and that the dynamical masses, with two exceptions, do not exceed $10^{10}~\msol$ \citep{maseda}.  Such a result would be entirely coincidental in the case that the emission lines are powered by AGN since the width of AGN emission lines is not coupled to the stellar mass of the host galaxy.  In other words, we observe narrow emission lines in these small ($\sim$ 1 kpc) systems, while typical AGN narrow line regions have much larget emission-line widths $\sigma > 200 \kms$ \citep{osterbrock}.  However, it is useful to look for evidence of AGN contributions.

\begin{figure}

\includegraphics[scale=.46]{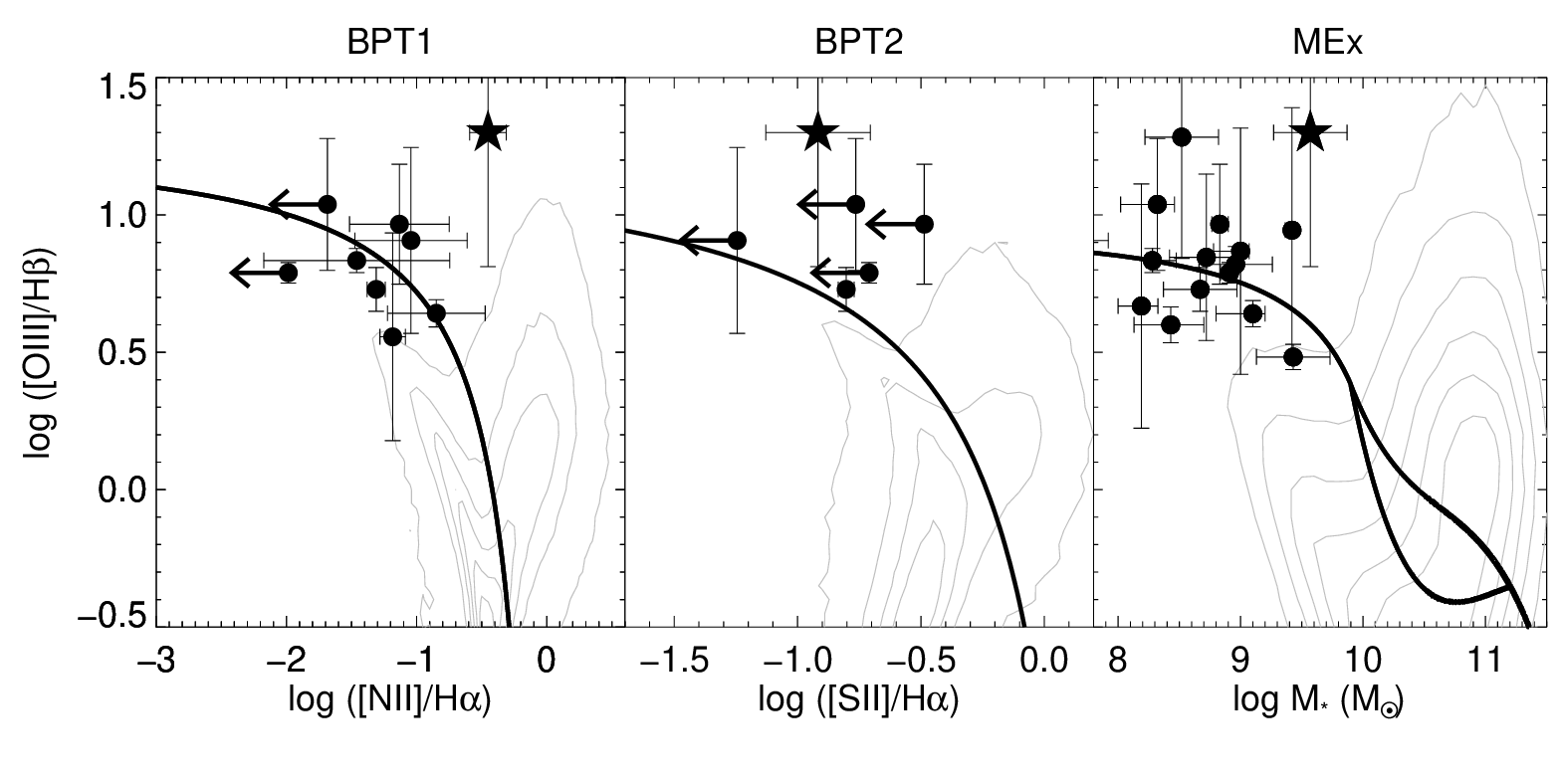}

\caption{AGN/SF emission line diagnostic plots.  From left to right, the BPT1 diagnostic of [N II]$\lambda$6584/H$\alpha$, the BPT2 diagnostic of [S II]$\lambda$6718+6731/H$\alpha$, and the MEx diagnostic \citep{juneau} of stellar mass.  Divisions between the AGN and the star-forming regions in the BPT diagrams come from \citet{kewley}.  In all cases, the gray contours represent data from the SDSS MPA-JHU value-added DR7 catalog: these emission line and stellar mass measurements are described by \citet{tremonti} and \citet{kauffmann}.  Arrows denote 3$\sigma$ upper limits.  Each of the diagnostics point to star formation as the ionizing source with at most mild AGN contribution, the possible exception being \textit{COSMOS-12102} (star symbol).  The large uncertainties in [O III]/H$\beta$ are caused by very low and uncertain H$\beta$ fluxes, as [O III] is robustly detected in all of these cases; the true ratio could be even higher than the values plotted here.}
\label{fig:mex}
\end{figure}

Although low-metallicity, low-mass AGN are exceedingly rare in the local Universe \citep{izotovAGN}, there is some evidence that they may be more common at higher-z \citep{trump11,xue,reines}. Such AGN could cause large observed line fluxes.  At $z> 1$, AGN identification with a simple diagnostic such a high [O III]$\lambda$5007/H$\beta$ ratio becomes insufficient by itself, as shown by \citet{trump}.  We thus utilize the ``BPT'' diagnostics of \citet{bpt} for the objects in our sample with measurements of [N II] and/or [S II] in addition to H$\alpha$.  Given that these lines are typically quite weak in star forming galaxies and the strong influence of OH skylines in our NIR spectra, in some instances we can only place an upper-limit on the ratios of those lines with H$\alpha$.  These two BPT diagrams are shown in the left and central panels of Figure \ref{fig:mex} with contours showing galaxies from the SDSS MPA-JHU value-added DR7 catalog.

Most of our points plausibly lie on an extension of the starforming region of the BPT diagrams and not on the extension of the AGN region, or at least they lie far from the main locus of AGN-powered emission lines.  However, low-metallicity AGN can lie in the starforming region as well \citep{groves,kewley13}, preventing us from completely ruling out the contribution of AGN to our sample with these diagnostics.  In some cases we observe higher [O III]/H$\beta$ ratios compared to the other starforming galaxies, but this can be explained simply by higher ionizations and lower metallicities in these systems compared to the low-z sample of SDSS galaxies.  \citet{kewley13} find that starforming galaxies at $z > 1$ are consistent with models that have more extreme ISM conditions than those in the local Universe\footnote{While beyond the scope of this paper, we would like to point out the large uncertainties in assumptions about the ISM conditions in galaxies at high redshift given the lack of knowledge about ionizing spectrum of hot stars at these early times and low metallicities.}.  High electron temperatures (discussed in 
the next section) support such a hypothesis.  This would be a further, unexplained coincidence if they are AGN in addition to
the low dynamical masses described previously.  While \citet{trump11} suggest that AGN are widespread in low-mass $1.3<z<2.4$ galaxies, the emission lines are not actually dominated by the AGN.  This is most evident in the $L_{0.5-10~keV}$/$L_{[O III]}$ relationship: the [O III] lines have some AGN contribution, but (on average) less than 50\%.

Additionally, we utilize the Mass-Excitation (MEx) diagnostic of \citet{juneau}, which combines the [O III]$\lambda$5007/H$\beta$ ratio with the stellar mass.  \citet{trump} conclude that this diagnostic also gives a meaningful probabilistic constraint on the AGN/SF nature of galaxies at $z > 1$ using a combination of the BPT diagnostics and X-ray data.  While this diagnostic is easily applicable to our data, we can not constrain the MEx AGN/starforming \textit{probabilities} for most of our sample given that it is not properly calibrated for objects with such low metallicities and high sSFRs, AGN or otherwise: the five objects with a probability of star formation ($P_{SF}$, compared to the probability that they are AGN; \textit{COSMOS-18358}, \textit{COSMOS-13184}, \textit{COSMOS-10599}, \textit{UDS-24154}, and \textit{UDS-12435}) have a median $P_{SF}$ value of 0.940, firmly placing them in the starforming regime.  The remaining objects, while probabilistically unconstrained, still lie far from the  
population of AGN in the 
MEx diagnostic plot, as shown in Figure \ref{fig:mex}.

We can place other constraints on the AGN nature of these systems in much the same way as \cite{vdw}, i.e. independent of any measured emission line ratio(s).  Most objects in our sample do not have strong 24 $\mu$m detections using Spitzer-MIPS: \textit{COSMOS-18358} has a MIPS 24$\mu$m flux of 21.5$\pm$9.0 $\mu$Jy and clearly appears to be a merger; \textit{COSMOS-19049} has a flux of 14.1$\pm$8.6 $\mu$Jy, is physically large with $r_{eff}=2.3$ kpc, and also has broad lines with $\sigma_{[O III]}=122 \kms$; \textit{COSMOS-12102} has a flux of 55.3$\pm$8.7 $\mu$Jy and is further discussed below.  While none of our objects have X-ray detections, GOODS-S is also the only field with sufficient depth to find $z\sim2$ AGN which are not quasars.  That being said, it is possible to hide even a rapidly accreting X-ray AGN in a low mass galaxy \citep{aird}.  The consistent (and resolved) $J_{F125W}$- and $H_{F160W}$-sizes, as well as the sizes of the emission lines in the grism spectra, rule out the presence of a strong point source dominating the emission.  

An exception could be \textit{COSMOS-12102}, illustrated by the star in Figure \ref{fig:mex}, which is also has the largest line width in our sample.  \citet{reines} find active black holes in similar mass dwarf galaxies with broad H$\alpha$ emission in the local Universe with $M_{BH} \lesssim 10^6~\msol$.  Using their relation (Equation 5) of $L_{H\alpha}$ and FWHM$_{H\alpha}$ to $M_{BH}$, \textit{COSMOS-12102} has a black hole mass of $\sim 10^{6.2}~\msol$, which is in their observed range.  The mass determination is fraught with systematic uncertainties, such as variations in the geometry of the broad-line region and that at least some of the H$\alpha$ luminosity comes from star formation, but we cannot conclusively rule-out some AGN contribution for this galaxy.

\subsection{Metallicity}
\label{sec:metals}
\begin{figure*}
\begin{center}
 \includegraphics[scale=.65]{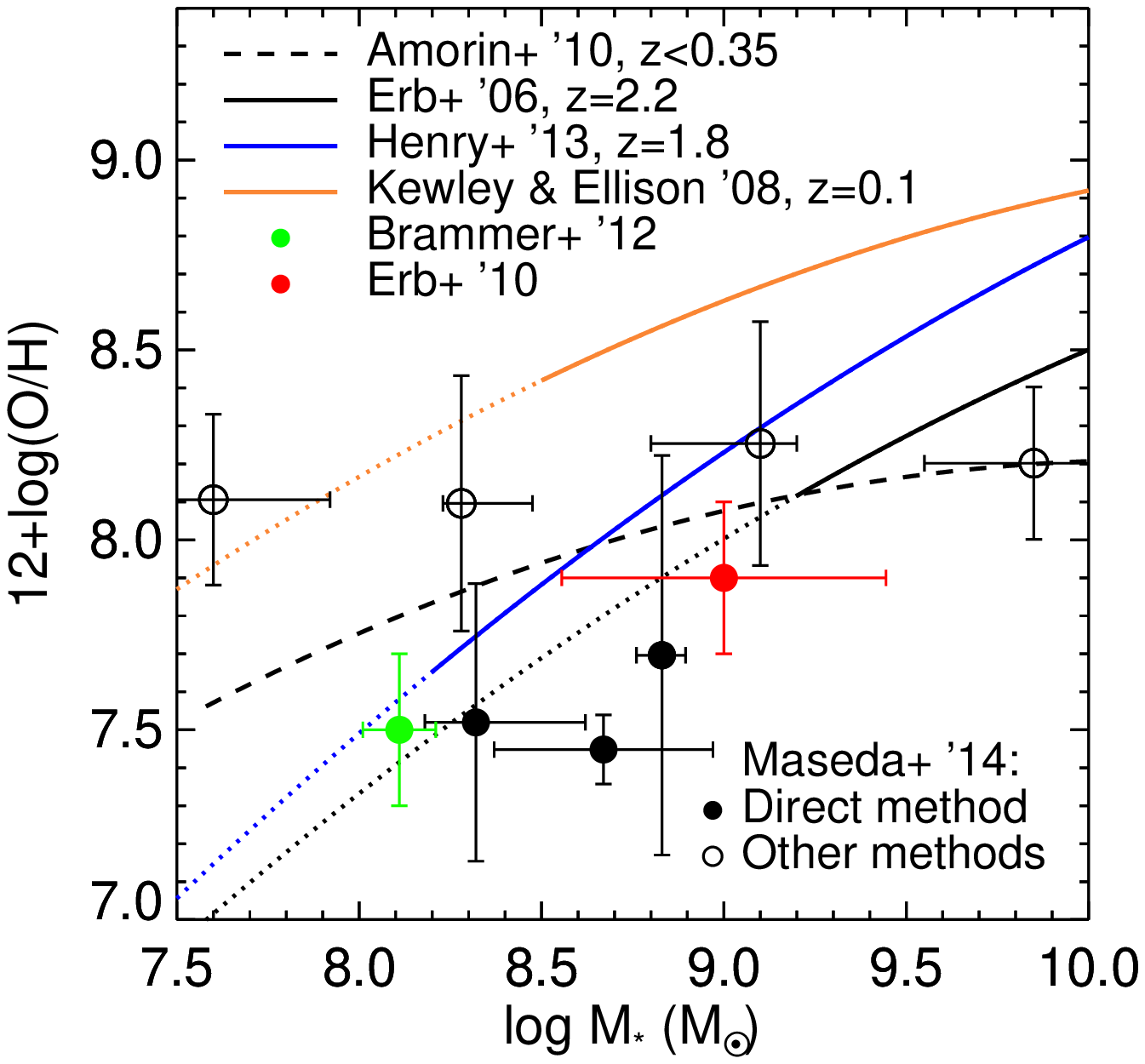}\includegraphics[scale=.65]{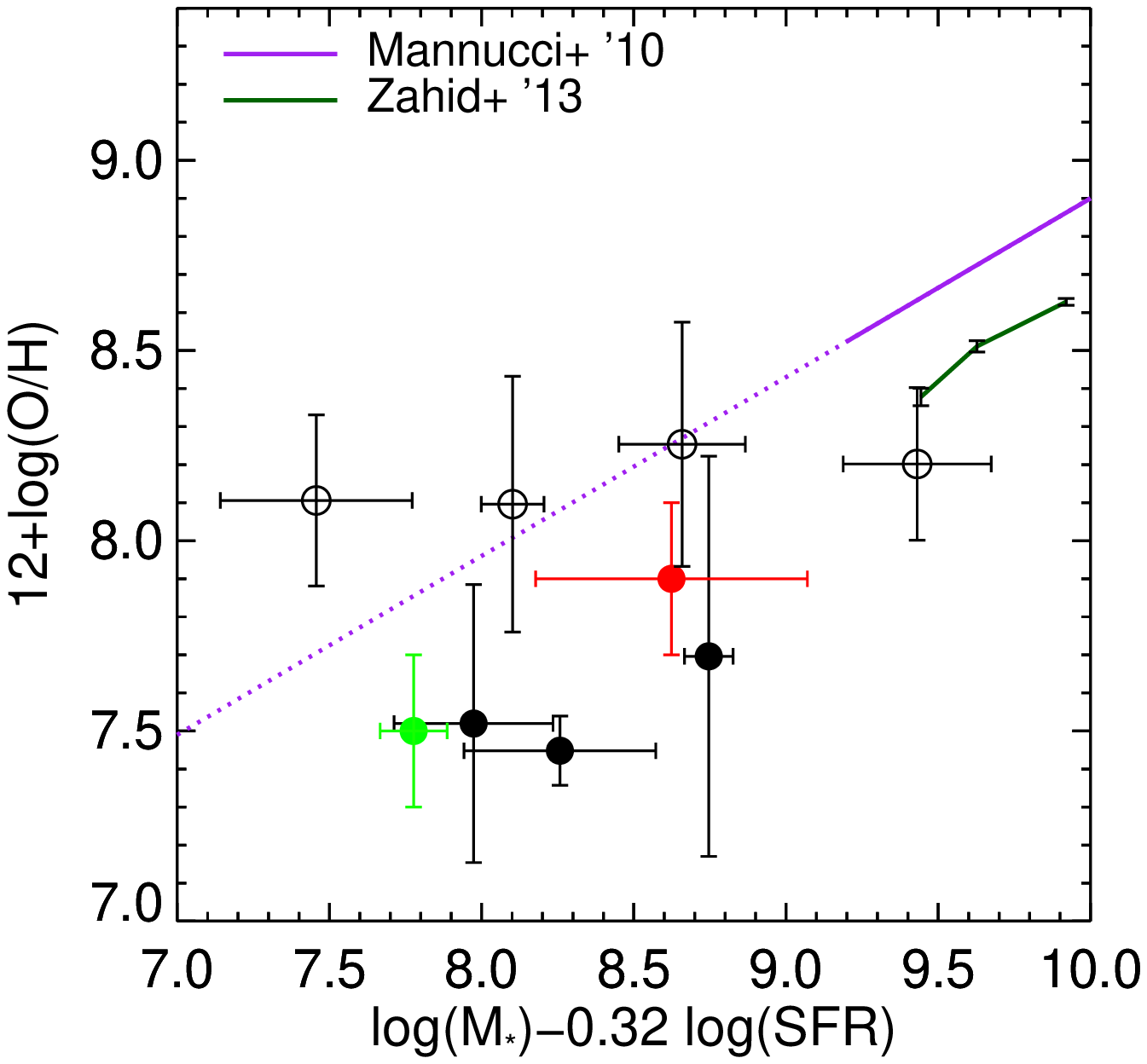}
\end{center}
\caption{Left panel: Oxygen abundances as a function of stellar mass. Overplotted are the MZ relations of \citet{kewleyellison}, \citet{henry13}, and \citet{erb06} with low-mass extensions given as the dotted lines, using the \citet{maiolino} parameterization.  The dashed line is the \citet{amorin} relation for luminous compact ``green pea'' galaxies at $0.11 < z < 0.35$. Right panel: Oxygen abundances for our sample as a function of $\mu_{32}$ as defined in \citet{mannucci}, with the ``FMRs" from \citet{mannucci} and the high-SFR bin from \citet{zahid} overplotted.  In both panels, filled points denote abundances obtained from the ``direct'' $T_e$ method and open points denote other methods.  The red point shows the result from \citet{erb10} and the green point shows the result from \citet{gb2}.  Overall we observe that our objects lie on or below the low-mass/high-SFR extrapolations to these observed relationships at similar redshifts.}  
\label{fig:metallicity}

\end{figure*}

In order to measure the gas-phase oxygen abundances of these galaxies as a proxy for the metallicity, we first implement the so-called ``direct'' or $T_e$ method which requires a detection of the auroral [O III]$\lambda$4363 line as well as the [O III]$\lambda\lambda$4959, 5007 and [O II]$\lambda$3727 lines.  Some of these lines lie in the UV/Visible at these redshifts, so we can only apply this method to the X-Shooter sample.  Using the calibrations outlined in \citet{izotov06}, we convert the [O III] emission-line ratios into an electron temperature (with the electron density constrained by the ratio of [O II]$\lambda$3729 to $\lambda$3726 or [S II]$\lambda$6717 to $\lambda$6731) in the O$^{++}$ region.  The total oxygen abundance in the galaxy is  O/H = O$^+$/H$^+$ + O$^{++}$/H$^+$, assuming $\log T_e($O$^+) = 0.7\log T_e($O$^{++})+0.3$.  Two objects, \textit{UDS-12539} and \textit{UDS-6377}, have detections of $\lambda$4363 and an upper limit can be obtained for a third, \textit{GOODS-S-43928}.  This object also lacks a detection of [O II]$
\lambda$3727 due to 
skyline 
contamination, so the O$^+$/H$^+$ component cannot be constrained directly.  The relative contribution of the O$^+$/H$^+$ component to the total oxygen abundance is 1.8\% and 7.3\% for the other two objects, so we neglect its contribution to the abundance of the third object.  The N2 method of \cite[][PP04]{pp04}, which uses the (log) ratio [N II]$\lambda$6584/H$\alpha$, verifies this result.  All derived temperatures are in excess of 20,000 K.

\begin{deluxetable}{lcc}
\tablecaption{Metallicity Estimates\label{tab:z}}
\tablehead{ \colhead{ID} & \colhead{12 + log(O/H)} & \colhead{Method}}
\startdata
GOODS-S-43693&8.10$\pm$0.34&N2\\
GOODS-S-43928&7.70$\pm$0.53&$T_e$/N2\\
UDS-6195&8.11$\pm$0.23&O3N2\\
UDS-6377&7.52$\pm$0.37&$T_e$\\
UDS-12539&7.45$\pm$0.09&$T_e$\\
UDS-24154&8.25$\pm$0.32&O3N2\\
COSMOS-19049&8.20$\pm$0.20&N2\enddata
\end{deluxetable}

For the remainder of the sample, we must resort to other methods, namely the aforementioned N2 method and also the O3N2 method, both from PP04; O3N2 is the (log) ratio [O III]$\lambda\lambda$4959, 5007/H$\beta$/N2.  For both methods, uncertainties include contributions from the error in the line ratios as well as the systematic scatter in the relations.  If H$\beta$ is not detected at more than 1$\sigma$ in our LUCI1 or X-Shooter spectrum, we estimate the [O III]/H$\beta$ ratio from the grism spectrum.  Also note that in the two instances where the N2 method is used the \citet{maiolino} calibration results in metallicities consistent with the PP04 values.  Results are displayed in Table \ref{tab:z}, with a median $12 + \log(O/H)$ value of 7.90 (0.15 $\zsol$).  This value agrees well with the median \texttt{MAGPHYS}-derived metallicity of 0.17 $\zsol$ for the full sample.

Given the relatively small number of direct measurements of the oxygen abundance in high-z galaxies, this provides an important piece of information.  Since the standard R23 diagnostic \citep{r23} using [O III]$\lambda\lambda$ 4959, 5007 + [O II] $\lambda\lambda$ 3726, 3729 +  H$\beta$ is double-valued, it is important whether higher-z galaxies belong to the metal-rich upper branch or to the metal-poor lower branch.  \citet{henry13a} argue in favor of the upper-branch for galaxies with log ($M_{\star}$/$\msol$) $>$ 8.2 at $z\sim0.7$, and \citet{henry13} favor the upper-branch for galaxies with log ($M_{\star}$/$\msol$) $>$ 8.8 at $1.3<z<2.3$.  Our measurements suggest that these systems plausibly lie on the lower branch, even at log ($M_{\star}$/$\msol$) $<$ 9.2.  

In the left panel of Figure \ref{fig:metallicity} we show these results as well as those of \citet{gb2} and \citet{erb10} with the low-mass extrapolations to the mass-metallicity (MZ) relations from \citet{kewleyellison} for $z \sim 0.1$, \citet{henry13} for emission line galaxies at $z \sim 1.8$, and \citet{erb06} for starforming galaxies at $z \sim 2.2$ utilizing the metallicity calibrations and functional parameterization of \citet{maiolino}, as well as the results from \citet{amorin}.  The right panel of Figure \ref{fig:metallicity} shows the same data plotted against the low-mass extrapolation of the ``fundamental metallicity relation'' (FMR) of \citet{mannucci} and the relation for high-SFR galaxies at $z\sim1.6$ from \citet{zahid}.  Neither method is directly calibrated below  $\sim 10^9~\msol$, which is precisely the range we probe here.

In the case of the MZ relation, our results generally agree with the \citet{erb06} and \citet{henry13} extrapolations and typically have lower metallicities than objects with similar masses and star formation rates at lower redshifts.  We see good agreement with distribution in $12 + \log(O/H)$ with $M_{\star}$ from \citet{amorin} as well.  The slight overall trend observed where higher-mass objects have higher metallicities supports the idea that the lowest-mass objects are the youngest and therefore have formed less mass of heavy elements.  The youngest objects should also have the highest sSFRs, which is known to drive much of the scatter in the MZ relation.  \citet{henry13a} also explain that the steepness of the O/H versus $M_{\star}$ relation has a direct theoretical impact 
on the role of outflows in these systems.

In the FMR relation the solid points, denoting $T_e$-determined abundances, are considered more reliable at low metallicities \citep{maiolino} and lie somewhat below the FMR, perhaps more in line with an extrapolation to the $z\sim1.6$ findings of \citet{zahid}.  Since \citet{mannucci} do not see the higher-mass turnoff in the relation until $z > 2.5$, we may be observing the evolution of the FMR on the low-mass and/or high-SFR end, although we do not have the number statistics yet to quantify any offset.  At the very least, we can conclusively rule-out that these objects lie above the FMR \cite[cf.][who claim this relation is driven by the higher average SFRs of the systems probed at higher redshifts]{stott}.

\section{Constraints on the Gas Fraction and the Star Formation Efficiency}
\label{sec:discussion}
In this section we use the observed velocity dispersions $\sigma$ and
sizes $r_{\rm{eff}}$, combined with dynamical stability criteria, to
constrain the gas fraction and its implications.  We assume that the
systems consist entirely of stars and gas: we neglect the contribution
of dark matter to the total dynamical mass as measured within the
central kpc.  We also assume that these systems are isolated and not embedded in larger (gaseous) structures that exert pressure.

We do not know the geometry of the systems, and therefore consider two
extreme cases: for the case of a sphere with uniform density we calculate the
Jeans mass $M_J$; for the case of a thin rotating disk we calculate
the Toomre parameter $Q$.  In both cases we assume that the gaseous
body has the same extent as the stellar body, and that the velocity
width of the nebular lines traces the total gas kinematics.

For a uniform sphere the Jeans mass \citep{binney} is given by
\begin{equation}
\label{eqn:jeans}
M_J = \frac{4\pi}{3} \rho_0\bigg(\frac{\pi\sigma^2}{4G\rho_0}\bigg)^{3/2},
\end{equation}
where we have equated the sound speed with observed velocity
dispersion.  This velocity represents the combined effect of all
sources of pressure that act against collapse, which include thermal
motions (associated with the physical sound speed) as well as
turbulence and streaming motions.

$\rho_0$ is the density of the gas, which is given by the gas mass
($f_{gas}\times M_{dyn}$) and the size:
\begin{equation}
\label{eqn:rho}
\rho_0=\frac{f_{gas}M_{dyn}}{(4/3)\pi r_{eff}^3},
\end{equation}
where $f_{gas}$ is the gas fraction.  The total dynamical mass
$M_{dyn}$ is taken from Equation \ref{eqn:dyn}, with a value of 5
 for the proportionality constant for consistency with
the case under consideration here: that of a sphere with uniform
density\footnote{Elsewhere in this paper we use a value of 3, which corresponds to other, more realistic geometries such as inclined disks and radially-concentrated density profiles (e.g., isothermal).}.

For typical values of $r_{eff}$ (1 kpc) and $\sigma$ (50 $\kms$)
we find that $M_J \simeq M_{gas} (\equiv f_{gas} M_{dyn})$ for
$f_{gas}=0.66$.  Given that substantial star formation in these
systems the gaseous body must be unstable: we conclude, assuming a
homogeneous gaseous sphere, that $f_{gas} \gtrsim 0.66$.

In order to address the question to what extent this conclusion is
affected by the chosen geometry, we now consider the other (opposite)
case, and assume that these systems are rotating disks, where instability can be described by the Toomre parameter $Q$
\citep{toomre,binney}
\begin{equation}
 Q_{gas} = \frac{\sigma_z \kappa}{\pi G \Sigma_{gas}}
\end{equation}
where $\sigma_z$ is the velocity dispersion perpendicular to the disk,
and $\Sigma_{gas}$ is the average gas-mass surface density, given by
\begin{equation}
\label{eqn:sigma}
 \Sigma_{gas} = \frac{M_{gas}}{2 \pi r_{eff}^2} = \frac{M_{dyn} f_{gas}}{2 \pi r_{eff}^2},
\end{equation}
where $f_{gas}$ is the gas fraction.  The epicyclic frequency, $\kappa$, in a rotating exponential disk is 
\begin{equation}
\label{eqn:kappa}
 \kappa = \sqrt{2}(v_t/r_{eff}),
\end{equation}
where $v_t$ is the circular velocity of the disk.

While $\sigma_z$ and $v_t$ are not observed directly, both contribute to the observed linewidth, $\sigma$.  The contribution of rotation to $\sigma$, assuming an average inclination of 60 degrees and an exponential disk, is $\sim v_{rot}/\sqrt{2}$, such that we have
\begin{equation}
\label{eqn:sigmaobs}
 \sigma_{obs}^2 = \sigma_z^2 + v_t^2/2,
\end{equation}
which is empirically supported \citep{kassin}.

Combining the above, we solve for $v_t$ as a function of $\sigma_{obs}$:
\begin{equation}
\label{eqn:vt}
 v_t^2 = \sigma_{obs}^2\left(1 \pm \sqrt{1-(9/4)f_{gas}^2Q_{gas}^2 } \right),
\end{equation}

such that an unstable system ($Q < 1$) requires
\begin{equation}
 f_{gas} > \frac{2}{3}
\end{equation}
in order to produce a unique, physical solution.

Remarkably, regardless of whether we assume a homogeneous sphere or a
rotating disk for the gas, we find that a high gas fraction is needed
to explain the observed star formation activity.  At the same time,
the non-negligible contribution of the stellar mass to the total mass
implies that $f_{gas}$ cannot be arbitrarily close to unity and should
be $\lesssim 0.9$ (see Figure \ref{fig:dynmass}).

A significant caveat is that 
the proportionality constants in Equations \ref{eqn:jeans},
\ref{eqn:rho}, \ref{eqn:dyn}, \ref{eqn:kappa}, \ref{eqn:sigmaobs}, and
\ref{eqn:vt} depend on the details of the assumed geometry and
dynamical structure; their variation can alter the threshold value of
$f_{gas}$.  In addition, we ignore the stabilizing effect of the
stellar disk on the gas disk, however this increases the implied gas
fraction further.

The median gas fraction in our sample is 72\% (i.e. the y-axis in
Figure \ref{fig:dynmass} is a proxy for $f_{gas}$), in agreement with
the theoretical calculation.  While some objects are observed to have
lower gas fractions, 18 out of our 22 objects are consistent with
$f_{gas} >$ 2/3 within 1$\sigma$.

\begin{figure}
\begin{center}
\includegraphics[width=0.475\textwidth]{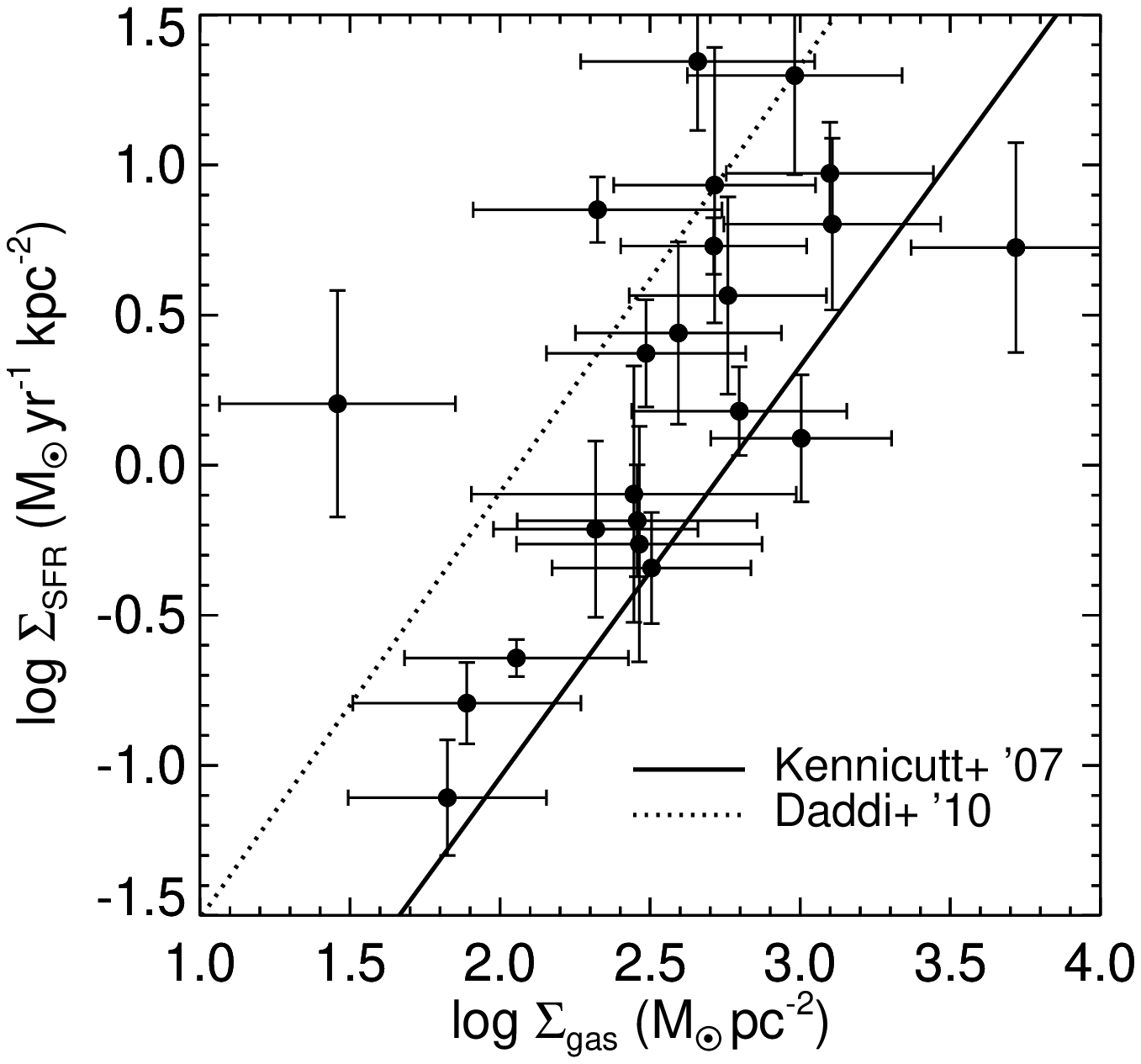}

\end{center}
\caption{SFR surface density versus gas-mass surface density.  The gas masses are estimated according to $M_{gas} = (2/3) \times M_{dyn}$, which is inferred as a probable gas fraction given the star formation rates and $M_{\star}$/$M_{dyn}$ ratios (which imply $0.5 \lesssim f_{gas} \lesssim 0.9$) in these systems.  The solid black line shows the relation for local spiral galaxies from \citet{kennicutt07} and the dotted line shows the result for local (U)LIRG and high-z SMGs/QSOs from \citet{daddi}.  The location of the points suggests that these objects spend at least some of the time forming stars more efficiently than the normal, present-day spiral galaxies.  Our constraints are not stringent enough to confirm or rule out gas depletion timescales that are on par with or even shorter than more massive, starbursting systems.}
\label{fig:gas}
\end{figure}

In the following we assume $f_{gas}$=2/3 in order to constrain the
star formation efficiency. In Figure \ref{fig:gas} we show the star
formation rate surface density, assuming $\Sigma_{SFR} = SFR / (2\pi
\times r_{eff})$, versus the gas surface mass density $\Sigma_{gas}$
(from Equation \ref{eqn:sigma}). The implied gas depletion time scale
ranges from $\tau_{depl} = 10^{7.5}$ to $10^9$ yr, with a median of $3
\times 10^8$ yr.  

Compared to normal present-day galaxies, gas is efficiently
transformed into stars but, with the exception of a few objects, not
as efficiently as observed in starbursting regions in the Milky Way
and starbursting galaxies in the present-day or high-redshift universe
\citep[i.e.][]{kennicutt07,daddi}.  This mostly reflects the large gas
fractions needed to produce systems that are unstable against star
formation, rather than a modest level of star formation: the inverse
sSFR (stellar mass growth time scale) of $1/sSFR = 5 \times 10^7$ yr
is very short, among the fastest ever measured.  Hence, the stellar
mass grows at a dynamical time scale ($\tau_{dyn}\sim3 \times 10^7$
yr).

The physical interpretation is that the star formation rate is not
limited by availability of fuel but by the dynamical time scale.
However, we have to keep in mind that we selected galaxies based on
their sSFR: we are biased against objects that are older and/or have
longer star formation time scales.  Further empirical investigation of
lower levels of star formation in similarly massive galaxies is needed
to address this issue.

However, we propose that star formation will halt within $\sim
50$~Myr, long before the gas reservoir is depleted.  First, the
stability arguments given above imply that the gas fraction only needs
to be reduced by a small amount (from the assumed $f_{gas}=2/3$ to,
say, $f_{gas}<0.5$).  Given the observed SFR, this takes ~$\sim$50
Myr.  Second, feedback should play an important role in these low-mass
systems with small escape velocities (several 100 $\kms$) and high
SFR: gas is easily transported out of the galaxy, and perhaps even out
of the halo, preventing recycling.  This paradigm is supported by \citet{law09,law12},
who see that higher mass starforming galaxies at z $\sim$ 2 can support more extended rotationally-supported disks
and are less efficient at driving outflows than their lower mass counterparts.

In the above we have ignored the infall of cold gas, which could
continue to feed and maintain the starburst.  Assuming that these
galaxies reside in relatively low-mass halos ($\sim10^{11} \msol$) the
typical accretion rate onto the halo is several $\msol$/yr
\citep{mcbride}, somewhat lower than the SFR.  The accretion rate onto
the halo is not necessarily equal to the accretion rate onto the
galaxy, and the latter is likely not constant.  A period of
above-average accretion for several 100 Myr could, in principle,
sustain the starburst.  However, above-average accretion events are
more likely to be of short duration, such that one such event can
ignite the observed starburst by pushing the gas mass surface density
above the threshold for star formation or by disturbing the
already-present gas, creating an instability.  Hence, enhanced accretion could cause the star formation
activity but not maintain it.

Based on the Jeans and Toomre instability arguments presented above,
purely gaseous systems with velocity dispersions and sizes as observed
become unstable once they reach a total mass of a few times
$10^9~\msol$, close to the observed masses of the objects in our
sample.

\section{Summary}
We presented near-infrared spectroscopy from the LBT/LUCI1
multi-object near-IR spectrograph and the VLT/X-Shooter wide band spectrograph
for a sample of HST/WFC3 grism-selected
emission line objects with restframe equivalent widths of $EW = 200 - 1100$ \AA$ $ for [O III]$\lambda$5007
and/or H$\alpha$, and located in the redshift range $1.3 < z < 2.3$. The observed emission lines are narrow,
with measured velocity dispersions down to $\sigma = 30$ km s$^{-1}$, implying low dynamical masses
of $\sim10^9~\msol$, even for the lower-EW objects not included in \cite{maseda}.  Stellar masses determined
using sophisticated \texttt{MAGPHYS} SED fitting to broadband magnitudes and the inclusion of line fluxes results in low stellar masses
as well, $\sim3\times10^{8}~\msol$.  Ratios of $M_{\star}$ to
$M_{dyn}$ range from 1/10 to 1, which makes AGN-dominated SEDs unlikely.  Emission-line ratios and the narrow line widths also suggest that AGN do not significantly contribute to our sample, and therefore we conclude that the main ionizing source is hot, massive stars.

Direct probes of the oxygen abundances within these galaxies and [O III]/H$\beta$ line ratios of typically $\gtrsim$ 5 corroborate the expectation that these low mass systems have low metallicities, between 0.05 and 0.3 $\zsol$.  They lie on or below the (extrapolated) mass-metallicity relationships for these redshifts \citep{henry13,erb06} which, combined with their young SED-derived ages, 
reinforces the notion that these are nascent galaxies undergoing their first major episode of star formation.  

Measured sSFR values of $\sim10^{-8}$ yr$^{-1}$ for these galaxies are up to two orders of magnitude larger than those of typical $10^{10}~\msol$ starforming galaxies at $z\gtrsim1$ \citep{fumagalli}, as well as comparable to or greater than the values from other high-EW systems as discovered in deep narrowband searches \citep{sobral} and in deep spectroscopic studies at both similar \citep{masters} and lower redshifts \citep{amorin14a,amorin14b}.  Such high sSFR values have been difficult to reproduce in hydrodynamical simulations, but recently \citet{shensims} made significant progress by combining a high gas density threshold for star formation and a blastwave scheme for supernova feedback in their simulations of low-mass galaxies.

Such low mass systems, with observed velocity dispersions of $\sigma\sim50\kms$ and sizes of $\sim$ 1 kpc are only unstable against star formation if their gas fractions are high (above 2/3), in agreement with the observed $M_{\star}/M_{dyn}$ relation.  The bursts are likely to be short-lived ($\sim$50 Myr), as, even in the absence of feedback, their intense star formation will rapidly build up stellar mass and lower their sSFR well before the gas depletion timescale ($\sim$100 Myr).

These results strengthen the conclusions from \citet{vdw}, who argued that EELGs represent
low-mass, starbursting galaxies. Additionally, the existence of (at least) two strong galaxy-galaxy lenses in the CANDELS/3D-HST fields where the background galaxy is an EELG at $z=1.85$ and $3.42$ \cite[][respectively]{gb2,vdw13} suggests that this type of object is common. The ubiquity of EELGs may be even more pronounced at high redshifts \cite[$>$6;][]{smit}.  Such systems at $z=$ 1 $-$ 2 thus may present an opportunity to study how star formation proceeded in the early Universe before the advent of the next generation of observatories, such as the \textit{James Webb Space Telescope}.

The new generation of submillimeter observatories, such as ALMA, can provide direct estimates of the gas masses. Searching for the presence of outflowing material would provide valuable clues about the feedback processes going on in these systems, which is especially relevant to support the hypothesis that these bursts can create the cored dark matter profiles observed in local dwarf galaxies \cite[e.g.][]{amorisco}.

\acknowledgements{M.V.M. is a member of the International Max Planck Research School for Astronomy
and Cosmic Physics at the University of Heidelberg, IMPRS-HD, Germany.  C.P. acknowledges support by the KASI-Yonsei Joint Research Program for the Frontiers of Astronomy and Space Science funded by the Korea Astronomy and Space Science Institute.  D.C.K. acknowledges funding from NSF grant AST-08-08133.}

{\it Facilities:} \facility{LBT}, \facility{VLT:Melipal}, \facility{HST}.
\bibliographystyle{apj}

\appendix
\begin{figure*}
 \includegraphics[width=.99\textwidth]{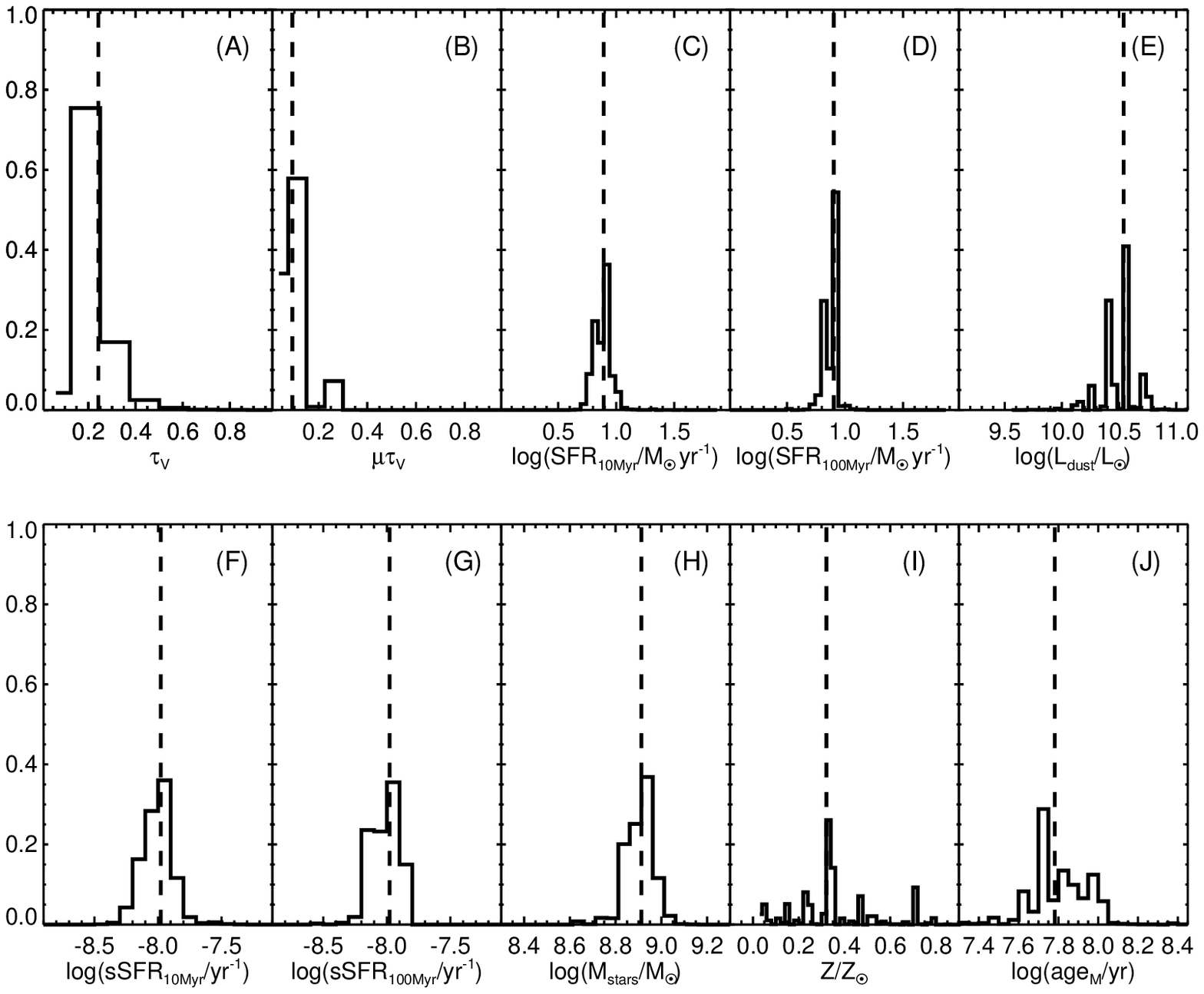}
 \caption{Probability distributions from \texttt{MAGPHYS} for \textit{GOODS-S-33131}.  Vertical dashed lines denote the medians of the output probability distribution, which are quoted throughout this work. Panels denote the following: (A) V-band optical depth seen by young stars in the birth clouds; (B) V-band optical depth seen by stars in the diffuse ISM; (C) star formation rate averaged over the last 10 Myr; (D) total dust luminosity; (E) star formation rate averaged over the last 10 Myr divided by stellar mass; (F) stellar mass; (G) stellar metallicity (which we set equal to the gas-phase metallicity); (H) mass-weighted age.}
 \label{fig:pdf}
\end{figure*}
\begin{figure*}
\begin{center}
 \includegraphics[width=.91\textwidth]{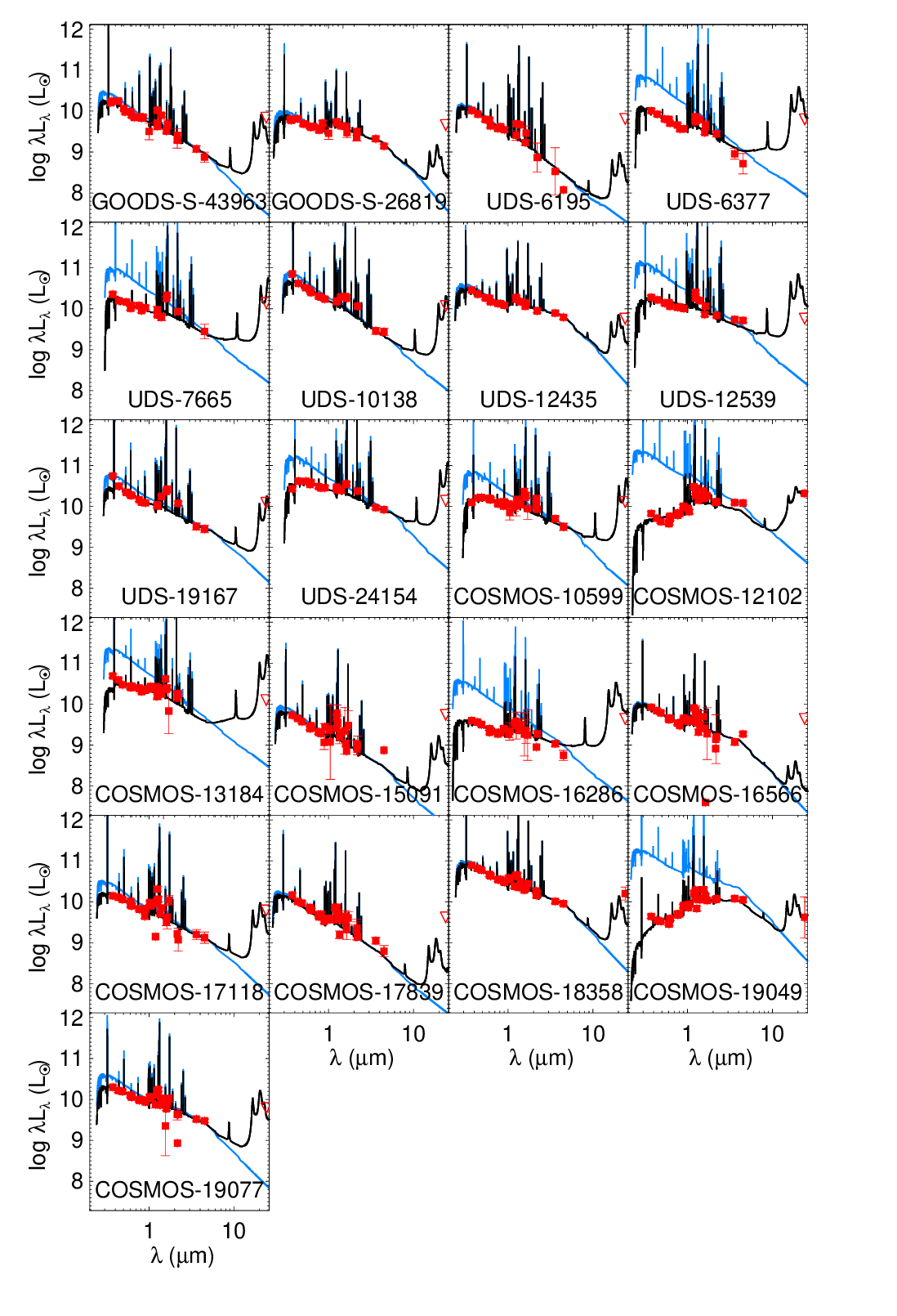}
 \end{center}
 \caption{Best-fit SEDs for the remaining objects in the sample.  The fits are performed as described in Section \ref{sec:sed}, with red points denoting the measured photometry (open points are upper limits), the blue curve denoting the total non-attenuated SED, and the black curve denoting the observed SED including dust attenuation.}
\label{fig:seds}
\end{figure*}

\end{document}